\documentclass[12pt]{article}
\usepackage{amscd,amsmath,amssymb,amsfonts,latexsym,mathrsfs,amsthm}
\usepackage{graphicx} 
\usepackage{setspace} 
\usepackage{graphics,epsfig}
\usepackage[english]{babel}
\usepackage{amsmath}
\usepackage{amsfonts}
\usepackage{amssymb,amsthm}
\usepackage{bm}
\usepackage{mathrsfs}
\usepackage{subfigure}
\usepackage[usenames]{color}
\usepackage{rotating}
\usepackage{url}
\usepackage{colortbl}
\usepackage{cite} 

\usepackage{hyperref}
\hypersetup{
  unicode=false,     
  pdftoolbar=true,    
  pdfmenubar=true,    
  pdffitwindow=false,   
  pdfstartview={FitH},  
  pdfauthor={},   
  pdfcreator={},  
  pdfproducer={}, 
  pdfkeywords={}{}, 
  pdfnewwindow=true,   
  colorlinks=true,    
  linkcolor={blue},     
  citecolor=blue,     
  filecolor=blue,     
  urlcolor=blue      
}

\definecolor{MYCOLOR0}{rgb}{0.92,0.92,0.92}

\newtheorem{theo}{Theorem}

\newtheorem{prop}{Proposition}

\usepackage{flushend}
\usepackage{verbatim}
\usepackage{tabularx}
\usepackage{multirow} 
\usepackage{arydshln}

\UseRawInputEncoding

\begin{document}

\title{Compressed particle methods for expensive models \\
with application in Astronomy and Remote Sensing}


\author{Luca Martino, V{\'i}ctor Elvira, Javier L\'opez-Santiago, Gustau Camps-Valls
\thanks{Research funded by the European Research Council (ERC) under the ERC-CoG-2014 SEDAL project (grant agreement 647423).}
\thanks{Luca Martino is with Dep. of Statistical Signal Processing, Universidad Rey Juan Carlos (URJC), Madrid, Spain. (luca.martino@urjc.es)} 
\thanks{Victor Elvira is with IMT Lille Douai, Cit{\'e} Scientifique, Rue Guglielmo Marconi, 20145, Villeneuve dAscq 59653, France.} 
\thanks{Javier Lopez-Santiago is with Dep. of Statistical Signal Processing, Universidad Carlos III de Madrid (UC3M), Madrid, Spain.} 
\thanks{G. Camps-Valls is with the Image Processing Laboratory, Universitat de Val\'{e}ncia, Spain. http://isp.uv.es/ (gustau.camps@uv.es)}%
}

\maketitle

\begin{abstract}
In many inference problems, the evaluation of complex and costly models is often required. In this context, Bayesian methods have become very popular in several fields over the last years, in order to obtain parameter inversion, model selection or uncertainty quantification. Bayesian inference requires the approximation of complicated integrals involving (often costly) posterior distributions. Generally, this approximation is obtained by means of Monte Carlo (MC) methods. In order to reduce the computational cost of the corresponding technique, surrogate models (also called emulators) are often employed. Another alternative approach is the so-called Approximate Bayesian Computation (ABC) scheme. ABC does not require the evaluation of the costly model but the ability to simulate artificial data according to that model. Moreover, in ABC, the choice of a  suitable distance between real and artificial data is also required. 
In this work, we introduce a novel approach where the expensive model is evaluated only in some well-chosen samples. The selection of these nodes is based on the so-called compressed Monte Carlo (CMC) scheme. We provide  theoretical results supporting the novel algorithms and give empirical evidence of the performance of the proposed method in several numerical experiments. Two of them are real-world applications in astronomy and satellite remote sensing. 


\end{abstract}

\section{Introduction}

In many areas of science and engineering, systems are analyzed by studying physical models and running computer simulations, which serve as convenient approximations to reality. Depending on the body of literature, they are known as physics-based, processed-oriented, mechanistic models, or simply simulators~\cite{Santer03,Wescott13}. Simulators and their corresponding surrogate models are ubiquitous in physics, brain, Earth, climate, and social sciences~\cite{Raissi17,sandberg2013feasibility,Verrelst2016,Majda14958,Young11}, but also in industrial environments for developing new manufactured products and infrastructures, to quantify performance of engineering systems, to understand and assess supply chains, or in robotics and vehicle design~\cite{conti2010bayesian,koziel2013physics,koziel2013multi,cutajar2019deep}. Model simulations are needed to understand system behaviour, but also to perform interventional and counterfactual studies. 

Since common forward models (simulators) are computationally costly, both running simulations or inverting them for parameter prediction becomes a big challenge. 
Machine learning models are widely used to learn both the forward and inverse functions, and nowadays they routinely replace complex models and sub-components to improve {\it scalability} and {\it mathematical tractability}. These models are commonly known as {\it emulators} and report excellent accuracy-speedup trade-offs compared to simulators, besides elegant ways to do uncertainty quantification, error propagation, and sensitivity analysis~\cite{OHagan2006}. 

Bayesian methods are often applied for parameter inversion, model selection or uncertainty quantification \cite{Liu04b,Robert04}. In their common implementation, these techniques require the evaluation of the possible complex and costly model. When the model is particularly expensive (or its pointwise evaluation is impossible), generally two approaches are employed. In the first one, the true model is replaced by a surrogate model (i.e., an emulator) that could be adaptively improved \cite{Busby09,Gorissen10,Wang11,SVENDSEN2020107103}. Then, Bayesian inference is carried out on the approximate and cheaper model. The second approach is the so-called {\it approximate Bayesian computation} (ABC) \cite{Beaumont10,Simola19,TURNER201269}. In the standard ABC scheme, model evaluation is substituted by evaluating a distance between the observed data and some artificial data generated according to the model. Therefore, ABC does not need to evaluate the model but to simulate artificial data from it. Different types of distances can be used. It is important to remark the choice of distance can be interpreted as a choice of an approximate observation model. 

In this work, we consider an alternative approach. The core idea is to reduce the number of true model evaluations by a suitable selection of the inputs where we evaluate the model. In this way, we can obtain a great reduction in the required computational time, at the expense of a slight increase of the estimation error. The key point is a proper selection of inputs where to evaluate the costly model. We present the novel approach in the context of particle filtering where the variables of interest can also vary with time. The method is based on a technique called compressed Monte Carlo (CMC) scheme, which summarizes the information contained in $N$ weighted Monte Carlo samples into $M< N$ weighted particles (called {\it summary particles}), based on
a stratification approach \cite{mcbookOwen,Robert04}. We aim at reducing the loss of information in terms of moment matching, in a similar fashion of the deterministic-based quadrature rules \cite{CKF09,Julier04,Huszar_2012,Sarkka13bo,CMC18,CMC_Inf_Scie,IGQ_2020}. We provide different theoretical results supporting the novel approach. 

We also give empirical evidence of the performance of the proposed method 
in four different numerical experiments, both over simulations and real challenging problems. 
In particular, we 
consider the problem of object detection (planets, satellites, etc.) in an $N$-body system observed from the Earth. The observation model is complex and costly, especially for some set of parameters (see Section \ref{Simu}). The second real model 
considers the inversion of a radiative transfer model (RTM) 
which encodes the energy transfer through the atmosphere. This model is used to understand and model vegetation, 
as well as to estimate the parameters that describe the status of the Earth from satellite observations by inversion. 

The paper is structured as follows. Section \ref{MC_Sect} describes the problem statement and recalls some background material. In Section \ref{SectCMC}, we introduce the CMC approach. In Section \ref{AnCMC}, we provide some theoretical results. In Section \ref{DistrSect}, we introduce the novel compressed particle filtering algorithms. Numerical simulations are given in Section \ref{Simu}. Finally, some conclusions are provided in Section \ref{ConSect}.

\section{Problem statement}\label{MC_Sect}
In many real-world applications, it is required to characterize the posterior probability density function (pdf) of a set of unknown parameters given the observed data.
More specifically, denoting the vector of unknowns as ${\bf x}=[x_1,\ldots,x_{d_X}]^{\top}\in \mathcal{D}\subseteq \mathbb{R}^{d_X}$ and the observed data as ${\bf y}\in \mathbb{R}^{d_Y}$, the pdf is defined as
\begin{equation}
	{\bar \pi}({\bf x}| {\bf y})
		= \frac{\ell({\bf y}|{\bf x}) g({\bf x})}{Z({\bf y})} \propto \ell({\bf y}|{\bf x}) g({\bf x}) = \pi({\bf x}|{\bf y}),
\label{eq_posterior}
\end{equation}
where ${\bar \pi}({\bf x}| {\bf y})$ and $\pi({\bf x}| {\bf y})$ denote the normalized and unnormalized posterior, respectively,  $\ell({\bf y}|{\bf x})$ is the likelihood function, $g({\bf x})$ is the prior pdf, and $Z({\bf y})$ is the normalization factor, which is usually called {\it marginal likelihood} or {\it Bayesian model evidence}.
Hereinafter, we will remove the dependence on ${\bf y}$ to simplify the notation.
%
A generic integral involving the density of the random variable ${\bf X} \sim {\bar \pi}({\bf x})=\frac{1}{Z} \pi({\bf x})$ is given by
\begin{align}
	I{(h)} \triangleq \mathbb{E}_{\bar \pi}[h({\bf X})]&= \int_{\mathcal{D}} h({\bf x}) \bar{\pi}({\bf x}) d{\bf x}  \nonumber\\
	&= \frac{1}{Z} \int_{\mathcal{D}} h({\bf x}) \pi({\bf x}) d{\bf x},
\label{eq_integral}
\end{align}
where $h({\bf x})$ is an integrable function of ${\bf x}$.\footnote{We assumed $h({\bf x}):\mathbb{R}^{d_X}\rightarrow \mathbb{R}$ and the integral $I(h)\in \mathbb{R}$ is a scalar value. However, more generally, we have ${\bf h}({\bf x}):\mathbb{R}^{d_X}\rightarrow \mathbb{R}^{\nu}$ and ${\bf I}({\bf h})\in \mathbb{R}^{\nu}$ where $\nu\geq1$. For simplicity, we keep the simpler notation with $\nu=1$.} For the sake of simplicity, we assume that the functions $h({\bf x})$ and $\bar{\pi}({\bf x})$ are continuous in $\mathcal{D}$, and the integrand function, $h({\bf x}) \bar{\pi}({\bf x})$, in Eq. \eqref{eq_integral} is integrable.  In many practical scenarios, we cannot obtain an analytical solution for \eqref{eq_integral}, and Monte Carlo methods are often applied.  More generally, we are interested in obtaining a particle approximation ${\widehat \pi}^{(N)}({\bf x})$ of the measure of ${\bar \pi}({\bf x})$, formed by a cloud of weighted samples  \cite{Liu04b}. See next section, for further details.  


\subsection{Importance Sampling (IS) approximations} \label{ISsect}
%
A well-known Monte Carlo approach is the importance sampling (IS) technique \cite{7974876,Robert04}. Let us consider $N$ samples $\{{\bf x}_n\}_{n=1}^N$ drawn from a proposal pdf, $q({\bf x})$, such that $q({\bf x})>0$ where ${\bar \pi}({\bf x})>0$.
We also assume that $q({\bf x})$ has heavier tails than the target, ${\bar \pi}({\bf x})$, since this assumption ensures that the resulting IS estimator has finite variance\cite{Liu04b,Robert04}. We assign a weight to each sample and then we normalize them as follows,
\begin{equation} 
	w_n= \frac{\pi({\bf x}_n)}{{q({\bf x}_n)}}, \qquad \qquad \bar{w}_n=\frac{w_i}{\sum_{j=1}^N w_j},
\label{is_weights_static}
\end{equation} 
with $n=1,\ldots,N$. Therefore, the moment of interest can be approximated as
\begin{eqnarray}
	\widehat{I}^{(N)}{(h)}&=& \frac{1}{N\widehat{Z}} \sum_{n=1}^N w_n h({\bf x}_n) \\
	&=& \sum_{n=1}^N \bar{w}_n h({\bf x}_n),
\label{eq_partial_estimator_static}
\end{eqnarray}
where $\widehat{Z}=\frac{1}{N}\sum_{n=1}^N w_n$ is a unbiased estimator of the marginal likelihood $Z=\int_{\mathcal{D}} \pi({\bf x}) d{\bf x}$, which is a useful quantity for model selection and hypothesis testing \cite{Robert04}. The particle approximation of the measure of ${\bar \pi}$ is given by
\begin{equation}
\label{EqPiMonteCarlo}
{\widehat \pi}^{(N)}({\bf x})= \sum_{n=1}^N \bar{w}_n \delta({\bf x}-{\bf x}_n),
\end{equation}
where $\delta({\bf x})$ is the Dirac delta function. 


\subsection{Particle filtering for state-space models}

In the sequential scenario, the inference problem often concerns a sequence of variables of interest ${\bf x}_{0:T}=[{\bf x}_0,{\bf x}_1,\ldots,{\bf x}_T]$ (a.k.a., {\it trajectory}) given a sequence of related observations ${\bf y}_{1:T}=[{\bf y}_1,{\bf y}_2,\ldots,{\bf y}_T]$, where $T$ represents the last time step. The corresponding state-space model is completely defined by an initial density $p({\bf x}_0)$, a transition density and the likelihood function), i.e., 
 \begin{equation}
\left\{
\begin{array}{l}
{\bf x}_{t}\sim p({\bf x}_{t}|{\bf x}_{t-1}), \\
 {\bf y}_t\sim p({\bf y}_{t}|{\bf x}_{t}),
\end{array}
\right.
\qquad t=1,\ldots, T.
\end{equation}
The complete posterior density is given by
\begin{equation}
\bar{\pi}({\bf x}_{0:T}|{\bf y}_{1:T})\propto p({\bf x}_0)\left[\prod_{t=1}^T p({\bf x}_{t}|{\bf x}_{t-1})\right] \left[\prod_{t=1}^T p({\bf y}_{t}|{\bf x}_{t})\right].
\end{equation}
Efficient Monte Carlo techniques for approximating the posterior $\bar{\pi}({\bf x}_{0:T}|{\bf y}_{1:T})$ are the so-called particle filtering algorithms. A particle filter (PF) combines the sequential importance sampling approach with resampling steps. A standard PF is detailed in Table \ref{PFtable}. The resampling steps are performed when an effective sample size (ESS) approximation $\widehat{ESS}$ in smaller than $\eta N$ where $\eta\in [0,1]$ \cite{ESSarxiv16}. Examples of ESS are
\begin{align}
\widehat{ESS}\big(\bar{w}_{t}^{(1:N)} \big)&=\frac{1}{\sum_{n=1}^N (\bar{w}_{t}^{(n)})^2}, \\
 \widehat{ESS}\big(\bar{w}_{t}^{(1:N)} \big)&=\frac{1}{\max_n \bar{w}_{t}^{(n)}}.
\end{align}
{\bf Computational cost.} Note that, at each iteration, we have $N$ evaluations of the likelihood function (i.e., the observation model). Hence, after $T$ iterations of the filter, we have $NT$ model evaluations. Furthermore, the resampling steps (when performed) are done over $N$ possible particles. The cost of the resampling grows with $N$. More precisely, the complexity of the resampling procedure is of $O(N)$ \cite{Bolic04}.

\begin{table}[!h]
{
\caption{\normalsize A standard Particle Filter}
\label{PFtable}
\vspace{-0.3cm}
\begin{tabular}{|p{0.95\columnwidth}|}
  \hline
{\bf Initialization:} Choose $N$, $\eta\in [0,1]$, $\bar{\bf x}_{0}^{(i)}$, with $i=1,\ldots,N$, and $\widehat{ESS}$ \cite{ESSarxiv16}. Set $w_{0}^{(n)}=1$ for all $n$.\\
\\
{\bf For $t=1,\ldots,T:$}
\begin{enumerate}
\item Draw ${\bf x}_t^{(i)} \sim p({\bf x}_t |\bar{\bf x}_{t-1}^{(i)})$, with $i=1,\ldots,N$.
\item Compute the $N$ weights
\begin{equation}
w_{t}^{(n)}=w_{t-1}^{(n)} p({\bf y}_t|{\bf x}_t^{(i)}), \quad m=1,\ldots,M.
\end{equation}
and normalized them $\bar{w}_{t}^{(n)}=\frac{w_{t}^{(n)}}{\sum_{k=1}^N w_{t}^{(k)}}$. 
\item if $\widehat{ESS}\big(\bar{w}_{t}^{(1:N)} \big)\leq \eta N$:
\begin{itemize}
\item Obtain $\{\bar{\bf x}_t^{(n)}\}_{n=1}^N$, by resampling $N$ times within $\{{\bf s}_m\}_{m=1}^M$ according to $\bar{w}_{t}^{(m)}$, with $m=1,\ldots,M$.
\item Set $\widehat{Z}_t=\frac{1}{N}\sum_{n=1}^N w_{t}^{(n)}$ (see \cite{GIS18,GISssp16}), and 
$$
w_{t}^{(1)}=\ldots=w_{t}^{(N)}=\widehat{Z}_t.
$$
\end{itemize}
\end{enumerate} 
{\bf Outputs:} Return $\{{\bf x}_t^{(n)},w_{t}^{(n)}\}_{n=1}^N$ for $t=1,\ldots,T$. \\ 
\hline 
\end{tabular}
}
\end{table}

\section{Compressed Monte Carlo (CMC)}
\label{SectCMC}


In this section, we describe a procedure for compressing the information contained in a set of $N$ weighted samples $\{{\bf x}_n,\bar{w}_n\}_{n=1}^N$ obtained via importance sampling, with a smaller amount $M\leq N$ of weighted samples $\{{\bf s}_m,\widehat{a}_m\}_{m=1}^M$.\footnote{The case $M=N$ corresponds to the uncompressed scenario.} The CMC approach is based on the so-called stratified sampling \cite{Ecuyer94,mcbookOwen}. The idea is to divide the support domain $\mathcal{D}$ of the random variable ${\bf X}$ into $M$ disjoint, mutually exclusive regions. More specifically, let us consider an integer $M \in \mathbb{N}^+$ with $M<N$, and a partition $\mathcal{P}=\{\mathcal{X}_1,\mathcal{X}_2, \ldots.,\mathcal{X}_M\}$ of the state space with $M$ disjoint subsets, 
\begin{gather}
\begin{split}
\label{PartitionEq}
&\mathcal{X}_1\cup\mathcal{X}_2\cup\ldots\cup\mathcal{X}_M=\mathcal{D},\\
&\mathcal{X}_i\cap\mathcal{X}_k =\emptyset, \qquad i\neq k, \quad \forall i,j\in\{1,\ldots,M\}. 
\end{split}
\end{gather}
 We assume that all $\mathcal{X}_m$ are convex sets. Now, let us consider $N$ weighted samples $\{{\bf x}_n,{\bar w}_n\}_{n=1}^N$. Given the partition in Eq. \eqref{PartitionEq}, i.e., $\mathcal{X}_1\cup\mathcal{X}_2\cup\ldots\cup\mathcal{X}_M=\mathcal{D}$ formed by convex, disjoint sub-regions $\mathcal{X}_m$, 
we denote the subset of the set of indices $\{1,\ldots,N\}$, 
$$
\mathcal{J}_m=\{\mbox{all } i\in\{1,\ldots,N\}: \quad {\bf x}_i \in \mathcal{X}_m\},
$$
which are associated to the samples in the $m$-th sub-region $\mathcal{X}_m$. The cardinality $|\mathcal{J}_m|$ denotes the number of samples in $\mathcal{X}_m$, and clearly we have $\sum_{m=1}^{M} |\mathcal{J}_m|=N$.
\subsection{CMC approximation}\label{CMCapprox}
We can summarize the information contained in the particle approximation ${\widehat \pi}^{(N)}({\bf x})$ 
of Eq. \eqref{EqPiMonteCarlo}, by  constructing an empirical stratified approximation based on $M$ weighted particles $\{{\bf s}_m,\widehat{a}_m\}_{m=1}^M$ (where the ${\bf s}_m$ are {\it summary particles}), i.e.,
\begin{eqnarray} 
\label{CMC_PI_Eq}
{\widetilde \pi}^{(M)}({\bf x})= \sum_{m=1}^M \widehat{a}_m \delta({\bf x}-{\bf s}_m), \quad {\bf s}_m\in \mathcal{X}_m,
\end{eqnarray}
 where 
 \begin{gather}
 \begin{split}
\widehat{a}_m \approx \mathbb{P}({\bf X}\in \mathcal{X}_m)&=\int_{\mathcal{X}_m} {\bar \pi}({\bf x}) d{\bf x}=\frac{1}{Z}\int_{\mathcal{X}_m} \pi({\bf x}) d{\bf x}.
\end{split}
\end{gather}
We here refer to $\widehat{a}_m$ as summary weights, and to ${\bf s}_m$ as summary particles.

\subsection{Summary weights}
 The weights $\widehat{a}_m$ can be obtained using the IS approximation $\widehat{\pi}^{(N)}$ with $N$ samples, i.e.,
\begin{eqnarray}
\widehat{a}_m&=&\int_{\mathcal{X}_m} \widehat{\pi}^{(N)}({\bf x}) d{\bf x}= \sum_{n=1}^N {\bar w}_n \int_{\mathcal{X}_m} \delta({\bf x}-{\bf x}_i) d{\bf x}, \nonumber \\
&=& \sum_{n\in \mathcal{J}_m} {\bar w}_n.
\end{eqnarray}
By defining
\begin{eqnarray}\label{Zcis}
\widehat{Z}_m=\frac{1}{N} \sum_{i\in \mathcal{J}_m} w_i, \qquad \widehat{Z}= \sum_{m=1}^M \widehat{Z}_m=\frac{1}{N} \sum_{n=1}^N w_n,
\end{eqnarray}
we can also obtain another expression $\widehat{a}_m$, i.e.,
\begin{eqnarray}\label{Zm_over_Z}
\widehat{a}_m=\frac{\widehat{Z}_m}{\widehat{Z}}
=\sum_{i\in \mathcal{J}_m} \frac{w_i}{\sum_{n=1}^N w_n}= \sum_{i\in \mathcal{J}_m} {\bar w}_i.
\end{eqnarray}
where $\widehat{Z}_m \approx \int_{\mathcal{X}_m} \pi({\bf x}) d{\bf x}$ and $\widehat{Z} \approx \int_{\mathcal{X}} \pi({\bf x}) d{\bf x}$. Note that
$\sum_{m=1}^M\widehat{a}_m=1$,
i.e., they are normalized. Due to Eq. \eqref{Zm_over_Z}, the unnormalized CMC weights are defined as $a_m=\widehat{Z}_m \propto \widehat{a}_m$. They play the same role of the unnormalized weights in IS, indeed the arithmetic mean  of $a_m$'s is an estimator of the marginal likelihood.
\subsection{Summary particles}\label{SummPart}
We consider different strategies for the selection of the summary particles ${\bf s}_m$. The first one is a stochastic approach based on the stratified sampling: each summary particle 
${\bf s}_m$
 is resampled within the set of samples ${\bf x}_i \in \mathcal{X}_m$, i.e., 
 $$
 {\bf s}_m\in\{{\bf x}_i, \mbox{ with } i \in \mathcal{J}_m\},
 $$
 according to the normalized weights,
\begin{equation}
\label{EqREVmagic}
{\bar w}_{m,i}=\frac{w_i}{\sum_{k\in \mathcal{J}_m} w_k}=\frac{\bar{w}_i}{\sum_{k\in \mathcal{J}_m}\bar{w}_k}=\frac{\bar{w}_i}{\widehat{a}_m}, \quad i \in \mathcal{J}_m.
\end{equation}
Namely, in that case,
\begin{equation}\label{EqREVmagic2}
{\bf s}_m\sim {\widehat \pi}_m({\bf x})=\sum_{i\in\mathcal{J}_m} {\bar w}_{m,i} \delta ({\bf x}-{\bf x}_i).
\end{equation}
Deterministic choices are also possible, for instance setting
\begin{equation}
\label{VDet}
{\bf s}_m=\sum_{j\in \mathcal{J}_m} {\bar w}_{m,j} {\bf x}_{j}, 
\end{equation}
or, if we are interested on the approximation of a specific integral involving a function $h$, we can set
\begin{equation}
\label{EqSpecH}
s_m=\sum_{j\in \mathcal{J}_m}  {\bar w}_{m,j} h({\bf x}_{j}). 
\end{equation}
These deterministic rules provide a good performance and enjoy interesting properties, as discussed in the next section.
Table \ref{CMCsumm} summarizes the main notation of the work.

 \begin{table}[!h]
\small
{ 
\caption{Summary of the main notations.}
\label{CMCsumm}
\begin{tabular}{|c|c||c|c|}
  \hline
 \multicolumn{2}{|c||}{IS} & \multicolumn{2}{c|}{CMC}  \\ 
  \hline
 $w_n$ & ${\bar w}_n$ & $a_m$ &  $\widehat{a}_m$  \\
\hline
\hline
   &  &    &  \\ 
 $\frac{\pi({\bf x}_n)}{{q({\bf x}_n)}}$ & $\frac{w_n}{\sum\limits_{i=1}^N w_i}$ &  $\widehat{Z}_m=\frac{1}{N}\sum_{i\in \mathcal{J}_m} w_i$ & $\sum\limits_{i\in \mathcal{J}_m} {\bar w}_i$= $\dfrac{\widehat{Z}_m}{\widehat{Z}}$ \\
 & & & \\
\hline 
 \multicolumn{2}{|c||}{$n=1,\ldots,N$} & \multicolumn{2}{c|}{$m=1,\ldots,M$}  \\
\hline
\hline
\multicolumn{4}{|c|}{ } \\
 \multicolumn{4}{|l|}{Marginal likelihood estimator: $ \widehat{Z} =\sum_{m=1}^M \widehat{Z}_m=\frac{1}{N}\sum_{n=1}^N w_n$.} \\ 
\multicolumn{4}{|c|}{} \\
\multicolumn{4}{|l|}{Partial normalized weights: ${\bar w}_{m,i}=\frac{w_i}{\sum_{k\in \mathcal{J}_m} w_k}=\frac{\bar{w}_i}{\widehat{a}_m}, \quad i \in \mathcal{J}_m$.
 } \\
 \multicolumn{4}{|c|}{} \\
 \multicolumn{4}{|l|}{ CMC estimator: $ \widetilde{I}^{(M)}{(h)}= \sum_{m=1}^M \widehat{a}_m h({\bf s}_m)$. } \\
 \multicolumn{4}{|c|}{} \\
\hline
\end{tabular}
}
\end{table}

\noindent
{\it Case of unweighted samples.} Let us consider that we have $N$ samples $\{{\bf x}_n\}_{n=1}^N$ generated by a direct sampling method \cite{MARTINO_book}, or an MCMC algorithm \cite{Robert04}. The CMC scheme works in the same manner by setting $\widehat{a}_m=\frac{|\mathcal{J}_m|}{N}$, that represents the ratio of samples within $\mathcal{X}_m$. Moreover, in this scenario, ${\bar w}_{m,i}=\frac{1}{|\mathcal{J}_m|}$ for all $i \in \mathcal{J}_m$. 
\newline
\newline
{\it Examples of partition rules.} Given the $N$ samples ${\bf x}_n=[x_{n,1},\ldots,x_{n,d_X}]^{\top}\in \mathcal{D}\subseteq \mathbb{R}^{d_X}$, with $n=1,\ldots,N$. Then, we list three practical choices from the simplest to the more sophisticated strategy:
\begin{itemize}
\item[{\bf P1}] Random grid, where each component of the elements of the grid is contained within the intervals $\min\limits_{n\in \{1,\ldots,N\}} x_{n,i}$ and $\max\limits_{n\in \{1,\ldots,N\}} x_{n,i}$, for each $i=1,\ldots,d_X$.
\item[{\bf P2}] Uniform deterministic grid, where each component of the elements of the grid is contained within the intervals $\min\limits_{n\in \{1,\ldots,N\}} x_{n,i}$ and $\max\limits_{n\in \{1,\ldots,N\}} x_{n,i}$, for each $i=1,\ldots,d_X$.
\item[{\bf P3}] Voronoi partition obtained by a clustering algorithm with $M$ clusters (e.g., the well-known $k$-means algorithm). 
\end{itemize}
The procedures above are just possible examples. Note that using a  particular partitioning procedure, we can obtain different performance of the resulting algorithms. However, in all the proposed schemes, the theoretical and practical benefits  can be observed even applying the simplest rule P1, as we show in the next section and in the numerical experiments (Sect. \ref{Simu}).  Finally, note that even the simple procedures P1 and P2  take into account the sample information for building the partition.

\section{Properties of CMC}
\label{AnCMC}
In this section, we discuss some theoretical properties of the CMC schemes. The corresponding proofs are given below or in the related appendix. 
\newline
\newline
\noindent
{\bf Definition.} 
 A partition procedure is called {\it proper} if, when $M=N$, then $|\mathcal{J}_m|=1$ (note that $m=n$ in this case). Namely, in the limit case of $M=N$, we consider all the MC samples as summary samples, ${\bf s}_i={\bf x}_i$ for $i=1,\ldots,N$. 

{\theo\label{Teo2} Let us consider a fixed set of weighted samples $\mathcal{S}=\{({\bf x}_n,w_n)\}_{n=1}^N$
 and a given partition $\mathcal{P}$ (obtained with a proper procedure). Considering the stochastic selection of ${\bf s}_m\sim {\widehat \pi}_m({\bf x})$ in Eq. \eqref{EqREVmagic2}, the CMC estimator
 \begin{eqnarray}
 \widetilde{I}^{(M)}{(h)}= \sum_{m=1}^M \widehat{a}_m h({\bf s}_m), \qquad {\bf s}_m\sim {\widehat \pi}_m({\bf x}),
\end{eqnarray}
 is an unbiased estimator of $\widehat{I}^{(N)}{(h)}$ in Eq. \eqref{eq_partial_estimator_static}, i.e.,
 \begin{eqnarray}
 \mathbb{E}[\widetilde{I}^{(M)}{(h)}|\mathcal{S}]=  \widehat{I}^{(N)}{(h)}= \sum_{n=1}^N \bar{w}_n h({\bf x}_n). 
\end{eqnarray}
Furthermore, if the partition rule is proper for $M=N$, the CMC estimator coincides with exactly $\widehat{I}^{(N)}{(h)}$. }
\newline
\newline
{\bf Proof:} See Appendix \ref{appProof1} for the proof. $\Box$

{\theo\label{Teo3} Let us consider a proper partition procedure. As $M\rightarrow N$ and $N\rightarrow \infty$, the consistency of CMC estimator is ensured.}
\newline
\newline
{\bf Proof:} The IS estimator is consistent as~~~$N~~\rightarrow~~\infty$ \cite{Robert04}. If the partition procedure if proper, for $M=N$ the CMC estimator coincides with the standard IS estimator, i.e., $\widetilde{I}^{(M)}{(h)}=\widehat{I}^{(N)}{(h)}$ (recall $M=N$). Hence, the corresponding CMC estimator is also consistent. $\Box$
 
{\prop\label{Teo4} The estimator of the marginal likelihood $\widehat{Z}=\frac{1}{N}\sum_{n=1}^N w_n$ is reconstructed with no loss by the CMC estimator ${\widetilde I}^{(M)}=\frac{1}{M} \sum_{m=1}^M a_m$, i.e.,
${\widetilde I}^{(M)}=\widehat{Z}$.
}
\newline
\newline
{\bf Proof:} Since $a_m= \widehat{Z}_m$ (see Table \ref{CMCsumm}), we have 
$$
{\widetilde I}^{(M)}=\frac{1}{M} \sum_{m=1}^M a_m=\frac{1}{M} \sum_{m=1}^M \widehat{Z}_m =\widehat{Z},
$$
 as shown in Eq. \eqref{Zcis}. $\Box$
\newline 
\newline 
If we are interested only in one specific integral $I{(h)}= \int_{\mathcal{D}} h({\bf x}) \bar{\pi}({\bf x}) d{\bf x}$, it is convenient to apply CMC with the following deterministic choice of the summary particles 
\begin{equation}
\label{EqSpecH_2}
s_m=\sum_{j\in \mathcal{J}_m} {\bar w}_{m,j} h({\bf x}_{j}),
\end{equation}
as highlighted by the theorem below.
{\theo\label{Teo1} If ${s}_m$ is chosen as in Eq. \eqref{EqSpecH_2}, for $m=1,\ldots,M$,\footnote{Note that in this case $s_m\in \mathbb{R}$ is a scalar value since, for simplicity, we have assumed $h({\bf x}):\mathbb{R}^{d_X}\rightarrow \mathbb{R}$, instead of the more general assumption ${\bf h}({\bf x}):\mathbb{R}^{d_X}\rightarrow \mathbb{R}^{s}$ with $s\geq1$. All the considerations are also valid for $s\geq1$.} and the linear mapping $f(x)=x$, we have $\widehat{I}^{(N)}(h)={\widetilde I}^{(M)}(f)$, i.e., we have a perfect reconstruction of the IS estimator.} 
\newline
\newline
{\bf Proof:} See Appendix \ref{App1} for the proof. $\Box$ 

\section{Compressed Particle Filtering}
\label{DistrSect}

In this section, we show how CMC can be employed for a performance improvement or a decrease of the computational cost of benchmark particle filtering (PF) algorithms. Let us recall the state-space model
\begin{equation}
\left\{
\begin{array}{l}
{\bf x}_{t}|{\bf x}_{t-1}\sim p({\bf x}_{t}|{\bf x}_{t-1}), \\
 {\bf y}_t|{\bf x}_{t} \sim p({\bf y}_{t}|{\bf x}_{t}),
\end{array}
\right.
\qquad t=1,\ldots, T,
\end{equation}
described by the propagation kernel, $p({\bf x}_{t}|{\bf x}_{t-1})$, and the likelihood function $p({\bf y}_{t}|{\bf x}_{t})$.
Below, we provide two novel PFs based on CMC. In the first one, called compressed bootstrap particle filter (CBPF) and given in Table \ref{CBPF_table}, based on the so-called bootstrap particle filter, where the resampling is applied at each iteration. In the second one, described in Table \ref{GCPFtable}, where the resampling is applied at each iteration when $\widehat{ESS}\leq \eta N$. We describe the benefits of both compressed particle filter (CPF) techniques.
\newline
 {\bf Benefit 1.} In both proposed methods, the compression is applied before the evaluation of the likelihood function $p({\bf y}_t|{\bf x}_t)$. The reduction in computational cost is twofold (as shown also in the next point). First of all, both algorithms require the evaluation of the likelihood function only $M<N$ times, at the summary particles ${\bf s}_m$. 
This is particularly convenient if the evaluation of the likelihood is costly due to the number of data, or to a complex measurement model. 
\newline 
 {\bf Benefit 2.} The resampling step is performed over $M$ weighted samples instead of $N$. This advantage can be found also in other filters proposed in the literature \cite{Kotecha03a,Kotecha03b}, which present lower complexity than standard particle filters (decreasing the cost of the resampling steps). Recall that the computational complexity of the resampling procedure is of $O(N)$ in a standard PF, whereas in CPF is  $O(M)$ with $M\leq N$ \cite{Bolic04}.
 \newline 
 {\bf Benefit 3.} Additionally, the application of CMC also helps to prevent the sample impoverishment caused by the resampling operation as also shown in a similar approach \cite{Li_2012}.
This is due to the fact that the summary particles contain also spatial information regarding the uncompressed particles $\{{\bf x}_t^{(n)}\}_{n=1}^N$ \cite{Li_2012}. Therefore, the resampling in CPFs takes into account both, the normalized weights and spatial information (not only the weights, as in resampling steps in standard particle filters without applying CMC). The results in Section \ref{NumExaqui} confirm that the application of CMC ensures a better approximation of the empirical measure defined by $N$ weighted samples.
\newline
In summary, CPFs are clearly cheaper and faster than  the corresponding classical particle filters. Note that the CMC weights ${\widehat a}_m$ are included in particle weights in Eq. \eqref{AquiAm}. The weighted summary particles $\{{\bf s}_m,{\widehat a}_m\}_{m=1}^M$ play a similar role than the sigma points in the unscented Kalman filter (UKF) \cite{Julier04,Sarkka13bo}.
\newline
\newline
{\bf Remark.} If the computational time is mainly specified by the likelihood evaluation, the CPFs provide better performance than the corresponding standard particle filtering schemes, for a fixed time budget.  
\newline
\newline
{\bf Limitations and further considerations.} Compared with a standard filter with $N$ particles, the CPFs provide some performance loss in terms of estimation error, since the compressed filters use less likelihood evaluation ($M<N$). However, for a fixed budget of evaluations of the likelihood function, the compressed filters provide the best results, as depicted in Figure \ref{FigEX_CPF_3}, which shows the benefits of the proposed compression procedure.
The use of the compressed filters is recommended only when the cost of the evaluation of likelihood function is significantly higher than the operations required in the compression (e.g., inequalities checking and sums). Therefore, the use of the CPF is required only when the evaluation of the model is costly.
It is also remarkable that the CPFs share some features and also present a robust behavior, similarly to the so-called  {\it Gaussian particle filters} \cite{Kotecha03a,Kotecha03b} and the {\it approximate-grid particle filters} \cite{Arulampalam02,Ristic04}. Another limitation of the proposed scheme is that  the CPFs  are  less suitable for parallelization than standard schemes.
Finally, the design of refined adaptive partitions and adaptation of the compression to more sophisticated filters, as the auxiliary particle filters \cite{Pitt01}, deserve and require additional future works.


\begin{table}[!h]
{
\caption{\normalsize The Compressed Bootstrap Particle Filter (CBPF)}
\label{CBPF_table}
\vspace{-0.3cm}
\begin{tabular}{|p{0.95\columnwidth}|}
  \hline
{\bf Initialization:} Choose $N$, $M<N$, and $\bar{\bf x}_{0}^{(i)}$, with $i=1,\ldots,N$.\\
\\
{\bf For $t=1,\ldots,T:$}
\begin{enumerate}
\item Draw ${\bf x}_t^{(i)} \sim p({\bf x}_t |\bar{\bf x}_{t-1}^{(i)})$, with $i=1,\ldots,N$.
\item Apply a CMC scheme to $\{{\bf x}_t^{(n)},\frac{1}{N}\}_{n=1}^N$ obtaining $\{{\bf s}_m,{\widehat a}_m\}_{m=1}^M$.
\item Compute the $M$ weights
\begin{equation}
\label{AquiAm_0}
w_{t}^{(m)}={\widehat a}_m p({\bf y}_t|{\bf s}_m), \quad m=1,\ldots,M.
\end{equation}
and normalized them $\bar{w}_{t}^{(m)}=\frac{w_{t}^{(m)}}{\sum_{k=1}^M w_{t}^{(k)}}$.

\item Obtain $\{\bar{\bf x}_t^{(n)}\}_{n=1}^N$, by resampling $N$ times within $\{{\bf s}_m\}_{m=1}^M$ according to $\bar{w}_{t}^{(m)}$, with $m=1,\ldots,M$.
\end{enumerate} \\ 
\hline 
\end{tabular}
}
\end{table}

\begin{table}[!h]
{
\caption{\normalsize Generic CPF}
\label{GCPFtable}
\vspace{-0.3cm}
\begin{tabular}{|p{0.95\columnwidth}|}
  \hline
{\bf Initialization:} Choose $M$ and $N$ such that $N$ is a multiple of $M$, i.e., 
$$
K=\frac{N}{M}\in \mathbb{N}^{+}.
$$
 Moreover, choose $\eta\in [0,1]$, $\bar{\bf x}_{0}^{(i)}$, with $i=1,\ldots,N$,  and an effective sample size approximation $\widehat{ESS}$ \cite{ESSarxiv16}. Set $\rho_{0}^{(i)}=\frac{1}{N}$ for all $i=1,\ldots,N$. \\
\\
{\bf For $t=1,\ldots,T:$}
\begin{enumerate}
\item Draw ${\bf x}_t^{(i)} \sim p({\bf x}_t |\bar{\bf x}_{t-1}^{(i)})$, with $i=1,\ldots,N$.
\item Apply a CMC scheme to $\{{\bf x}_t^{(n)},\rho_{t-1}^{(n)}\}_{n=1}^N$ obtaining $\{{\bf s}_m,{\widehat a}_m\}_{m=1}^M$.
\item Compute the $M$ weights
\begin{equation}
\label{AquiAm}
w_{t}^{(m)}={\widehat a}_m p({\bf y}_t|{\bf s}_m), \quad m=1,\ldots,M.
\end{equation}
and normalized them $\bar{w}_{t}^{(m)}=\frac{w_{t}^{(m)}}{\sum_{k=1}^M w_{t}^{(k)}}$. 
\item if $\widehat{ESS}\big(\bar{w}_{t}^{(1:M)} \big)\leq \eta M$:
\begin{itemize}
\item Obtain $\{\bar{\bf x}_t^{(n)}\}_{n=1}^N$, by resampling $N$ times within $\{{\bf s}_m\}_{m=1}^M$ according to $\bar{w}_{t}^{(m)}$, with $m=1,\ldots,M$.
\item Set 
$$
\rho_{t}^{(1)}=\ldots..=\rho_{t}^{(N)}=\sum_{m=1}^M w_{t}^{(m)}.
$$ 
For further details see \cite{GIS18,GISssp16}.
\end{itemize}
\item Otherwise, if $\widehat{ESS}\big(\bar{w}_{t}^{(1:M)} \big)>\eta M$, set
$$
\bar{\bf x}_{t}^{(n)}={\bf s}_m, \quad \rho_{t}^{(n)}=w_{t}^{(m)},\mbox{  } \mbox{ with }  \mbox{  }  m= \Big\lceil \frac{n}{K} \Big\rceil,
$$
with $n=1,\ldots,N$.
\end{enumerate} \\ 
\hline 
\end{tabular}
}
\end{table}

\noindent
{\it Regularized Resampling.} In the resampling steps of CPFs, we have
\begin{equation}\label{ResEq}
\bar{\bf x}_t^{(n)}\sim \sum_{m=1}^M \bar{w}_{t}^{(m)}\delta({\bf x}-{\bf s}_m).
\end{equation}
The use of CMC provides clear advantages as discussed above in  Benefit 3.
 Additionally, in both CPF schemes, we can also employed a regularized resampling in order to reduce also the loss of diversity in the cloud of particles. 
 We can replace the delta functions in Eq. \eqref{ResEq} with other kernel functions. For instance, we can consider Gaussian kernels $K({\bf x}|{\bf s}_m,{\bm \Sigma}_m)$, of mean ${\bf s}_m$ and with a $d_X\times d_X$ covariance matrix ${\bm \Sigma}_m$ the $d_X\times d_X$ obtained by an empirical estimation considering the samples in $\mathcal{X}_m$, i.e., 
\begin{eqnarray}
\label{Sigma_m}
{{\bm \Sigma}_m=\sum_{j\in \mathcal{J}_m} {\bar w}_{m,j} ({\bf x}_j-{\bf s}_m) ({\bf x}_j-{\bf s}_m)^{\top}+\epsilon {\bf I}},
\end{eqnarray}
where ${\bf s}_m$ is defined in Eq. \eqref{VDet} and $\epsilon>0$. Hence, in this case, we have
\begin{eqnarray}
\label{RegResEq}
\bar{\bf x}_t^{(n)}\sim \sum_{m=1}^M \bar{w}_{t}^{(m)} K({\bf x}|{\bf s}_m,{\bm \Sigma}_m),
\end{eqnarray}
where $K(\cdot)$ represents a kernel function with location parameter ${\bf s}_m$ and covariance matrix ${\bm \Sigma}_m$. A similar regularized resampling is implicitly used in \cite{Kotecha03a,Kotecha03b}.
\newline
\newline
{\it Adapting $M$.} Let us consider the CBPF algorithm in Table \ref{CBPF_table}. We can adapt the number of summary particles $M$ used at each iteration. 
The underlying idea is that if $\widehat{ESS}$ is small, we need a less number $M$ of summary particles to summarize the information in $\{{\bf x}_n,\bar{w}_n\}_{n=1}^{N}$. Otherwise, if $\widehat{ESS}$ is high, we could need more summary particles for encoding all the statistical information contained in $\{{\bf x}_n,\bar{w}_n\}_{n=1}^{N}$. Then we can set
\begin{equation}
M_t=\max\left[\gamma\big\lfloor\widehat{ESS}\big\rfloor, M_{\texttt{min}}\right] \quad t=1,\ldots,T,
\end{equation}
with $\gamma>0$ and $M_{\texttt{min}}\geq 1$.



\section{Numerical experiments}
\label{Simu}

In the section, we test the proposed method in five different numerical experiments, comparing its performance with benchmark methods. The first three numerical examples consider artificial models and simple distributions, showing the advantages of the proposed compression scheme even in these scenarios. The fourth numerical experiment considers the problem of object detection (planet, satellite, etc.) in an $N$-body system observed from the Earth. The observation model is complex and costly, especially for some set of parameters. In the last experiment, we consider the inversion of a radiative transfer model (RTM) called PROSAIL, which models the energy transfer through the atmosphere. This model is used to model and understand vegetation status 
from satellite observations. 


\subsection{CMC versus standard resampling}\label{NumExaqui} 

For simplicity, Let us consider $x\in \mathbb{R}^{+}$. Moreover, we consider two possible target densities: the first one is a Gamma pdf
\begin{equation}
\bar{\pi}(x)\propto x^{\alpha-1}\exp\left(-\frac{x}{\kappa}\right),
\end{equation}
with $\alpha=4$ and $\kappa=0.5$, and the second one is a mixture of two Gaussians, with $x\in \mathbb{R}$,
\begin{equation}
\bar{\pi}(x)= \frac{1}{2}\mathcal{N}(x|-2,1)+\frac{1}{2}\mathcal{N}(x|4,0.25).
\end{equation}
{\it Experiment:} At each run, we generate $N=10^5$ Monte Carlo samples $\{x_n,\frac{1}{N}\}_{n=1}^N$ from the target pdfs. 
We compare the standard resampling (SR) strategy with different CMC schemes. Namely, with SR, we resample uniformly $M$ times within $\{x_n\}_{n=1}^N$ obtaining $\{s_m,\frac{1}{M}\}_{m=1}^M$ and, with the CMC schemes, we obtain $\{s_m,\widehat{a}_m\}_{m=1}^M$. Then, at each run, we compute the Root Mean Square Error (RMSE) for estimating the first $5$ moments of the corresponding target pdf (using the $M$ summary particles). Regarding the CMC schemes, we consider two kind of partition procedures: random (P1) and uniform (P2) described in Section \ref{CMCapprox}. Furthermore we compare the stochastic and the deterministic choices of the summary particles $s_m$ described in Section \ref{SummPart}. For the deterministic CMC we refer to the use of Eq. \eqref{VDet} for $s_m$. We repeat the experiment $10^3$ independent runs and average the results.
 
\begin{figure}[htbp]
\centering
\subfigure[Gamma target pdf]{\includegraphics[width=8cm]{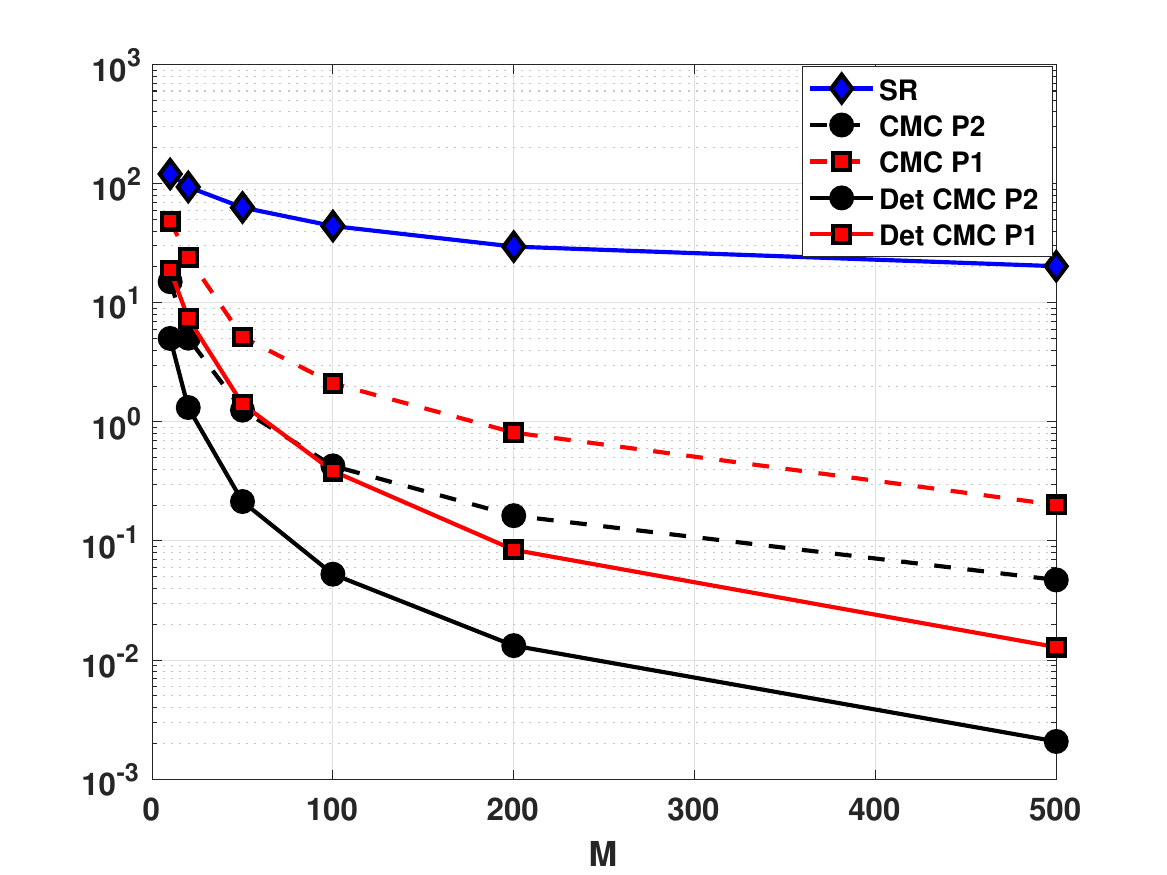}}
\subfigure[Mixture target pdf]{\includegraphics[width=8cm]{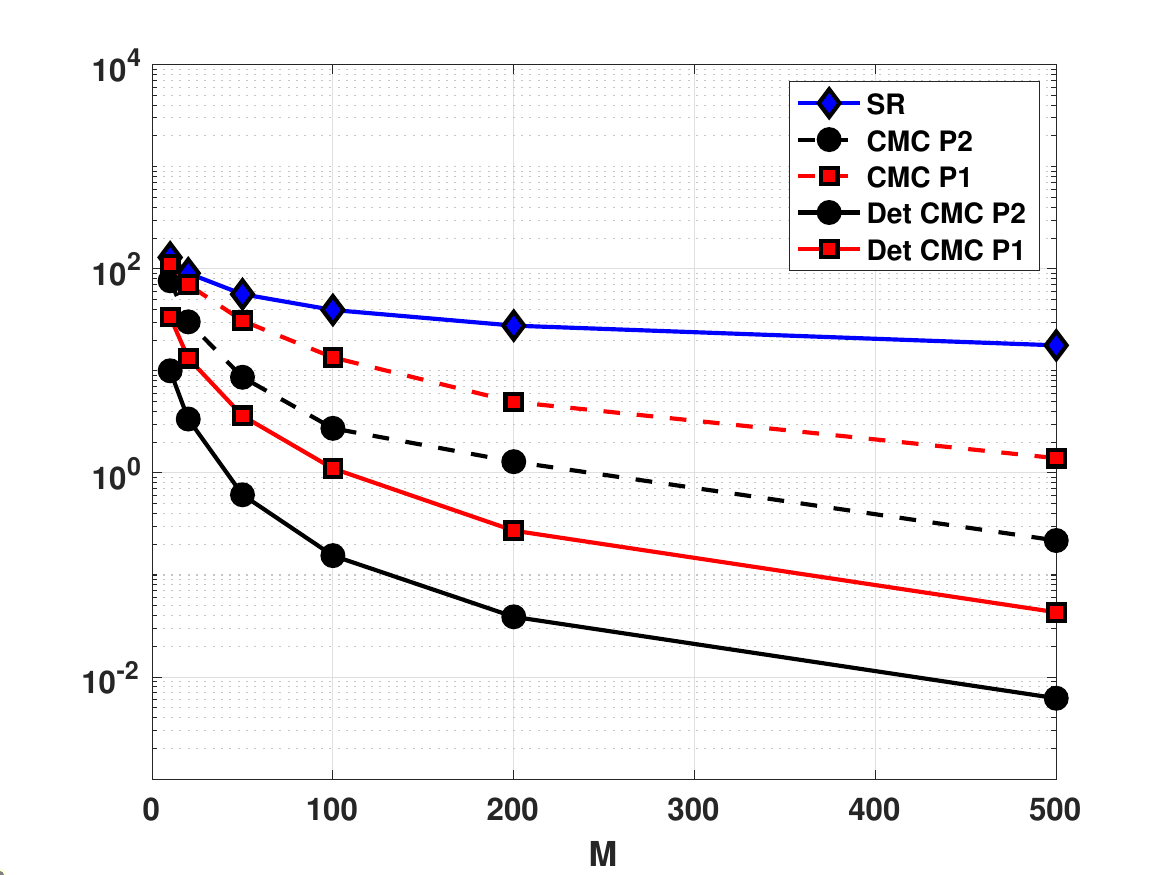}}
\caption{RMSE as function of $M$. The results obtained by SR is depicted with a solid line and rhombuses. The results of CMC with a random partition (P1) and with a grid partition (P2) are shown by squares and circles, respectively. The results obtained with the deterministic choice of ${\bf s}_m$ in Eq. \eqref{VDet} are shown with solid lines (squares and circles), whereas the results random choice of ${\bf s}_m$ are provided with dashed lines (squares and circles). }
\label{FigEX1}
\end{figure}

 Figure \ref{FigEX1} depicts the averaged RMSE as function of the number $M$ of summary particles.
 Figure \ref{FigEX1}-{\bf (a)} refers to the Gamma target pdf, whereas Figure \ref{FigEX1}-{\bf (b)} corresponds to the Gaussian mixture pdf. The results of the SR method are displayed with triangles. The stochastic CMC schemes are shown with dashed lines, whereas the deterministic CMC schemes with solid lines. 
\newline
{\it Discussion:} In all cases, CMC outperforms SR and the deterministic CMC schemes provide the better results. Clearly, the partition P2 (circles) outperforms P1 (squares). Note that P1 represents the simplest and perhaps the worst possible construction of the partition. However, it is important to remark that the CMC schemes, even with P1, outperform the SR method. In this experiment, the differences in computational time are negligible, and the CMC schemes provide always the best performance.

\subsection{Second Experiment}
\label{CPFsimu}
This section is devoted to analyze the performance of the compressed bootstrap particle filter (CBPF) described in Table \ref{CBPF_table}.
Let us consider the state-space model 
\begin{equation}
\left\{
\begin{array}{l}
x_{t}=| x_{t-1}|+ v_{t}, \\
 y_t=\log(x_{t}^2)+ u_{t},
\end{array}
\right.
\qquad t=1,\ldots, T,
\end{equation}
where $v_{t}\sim \mathcal{N}(0, 1)$ and $u_{t}\sim \mathcal{N}(0, 1)$. The goal is to track $x_{t}$ for $T=100$ steps, with a particle filtering algorithm considering $N\in\{100, 1000\}$ particles.
We compare the bootstrap particle filter (BPF) \cite{SMC01} with its compressed version (i.e., CBPF) in terms of the Root Mean Square Error (RMSE) in estimation of $x_{1:T}$. We apply CBPF with different values of $M$ (clearly, with $M\leq N$). 
We consider the deterministic CMC scheme with a uniform construction P2 of the partition.

\begin{figure}[htbp]
\centering
\includegraphics[width=8cm]{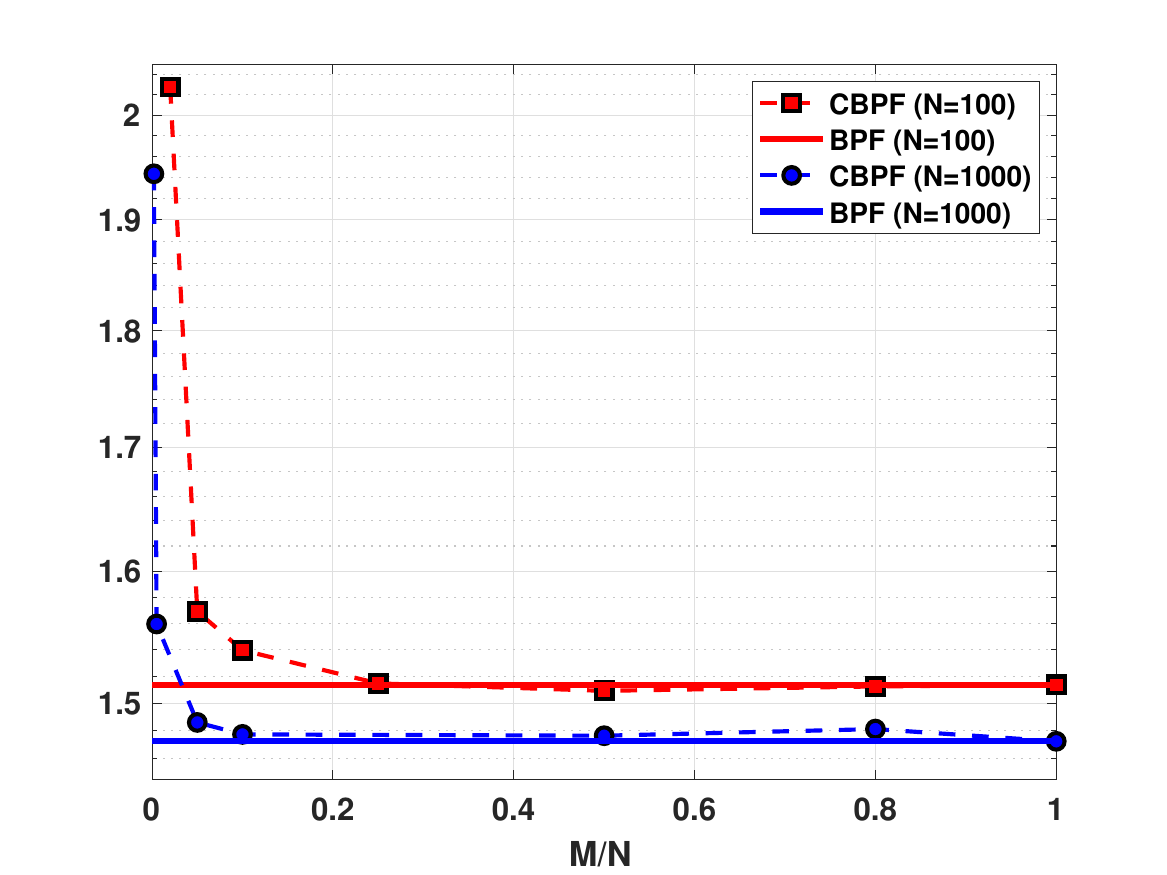}
\caption{RMSE (log-domain in the $y$-axis) as function of the ratio $\frac{M}{N}$. The dashed line with squares corresponds to the CBPF with $N=100$, whereas with circles corresponds to the CBPF with $N=1000$. The solid lines corresponds to the bootstrap particle filter (BPF) with $N=100,1000$. CBPF virtually obtains the same performance of the bootstrap particle filter with approximately $85\%$ less evaluations of the likelihood function.}
\label{FigEX_CPF}
\end{figure}

Figure \ref{FigEX_CPF} shows the RMSE (averaged over $5000$ independent runs) as function of the compression rate $\frac{M}{N}$. The solid lines represent the RMSE obtained by the BPF. The dashed line with squares corresponds to the CBPF (using the deterministic compression) with $N=100$, whereas the dashed line with circles corresponds to the CBPF with $N=1000$. Note that CBPF virtually obtains the same performance of the BPF with approximately $85\%$ less evaluations of the likelihood function. Clearly, in this toy example, the likelihood evaluation is not expensive, and the differences in computational time are negligible. This is not the case in the two real-world experiments, in Sections \ref{KeplerSect} and \ref{RemoteSect}, where the gain in computational time is relevant.
 We also recall that the $N$ resampling steps are performing over $M$ particles instead of $N$. Furthermore, fixing the compression rate $\frac{M}{N}$, It is interesting to note that the performance of CBPF improves when $N$ grows. 
 
\subsection{Third Experiment}
We now repeat the previous experiment considering another state-space model. More specifically, we consider the benchmark  growth model, as in \cite{Arulampalam02}, using also the  the same parameters as in \cite{Arulampalam02}, i.e.,
\begin{equation}
\left\{
\begin{array}{l}
x_{t}=f_t(x_{t-1})+ v_{t}, \\
 y_t=\frac{1}{20}x_{t}^2+ u_{t},
\end{array}
\right.
\qquad t=1,\ldots, T,
\end{equation}
where 
$$
f_t(x_{t-1})=\frac{1}{2}x_{t-1}^2+\frac{25x_{t-1}}{1+x_{t-1}^2}+ \cos(1.2 t),
$$
and  $v_{t}\sim \mathcal{N}(0, 10)$ and $u_{t}\sim \mathcal{N}(0, 1)$.  The goal is to estimate the temporal trajectory of the state $x_{t}$ for $T=100$ steps, with a particle filtering algorithm considering $N\in\{100, 1000\}$ particles.
Again, we compare the bootstrap particle filter (BPF) \cite{SMC01}, with its compressed version, CBPF, with different values of $M\leq N$. in terms of the Root Mean Square Error (RMSE) in the estimation of the trajectory $x_{1:T}$.  We consider the deterministic CMC scheme with a uniform construction P2 of the partition.

\begin{figure}[htbp]
\centering
\includegraphics[width=8cm]{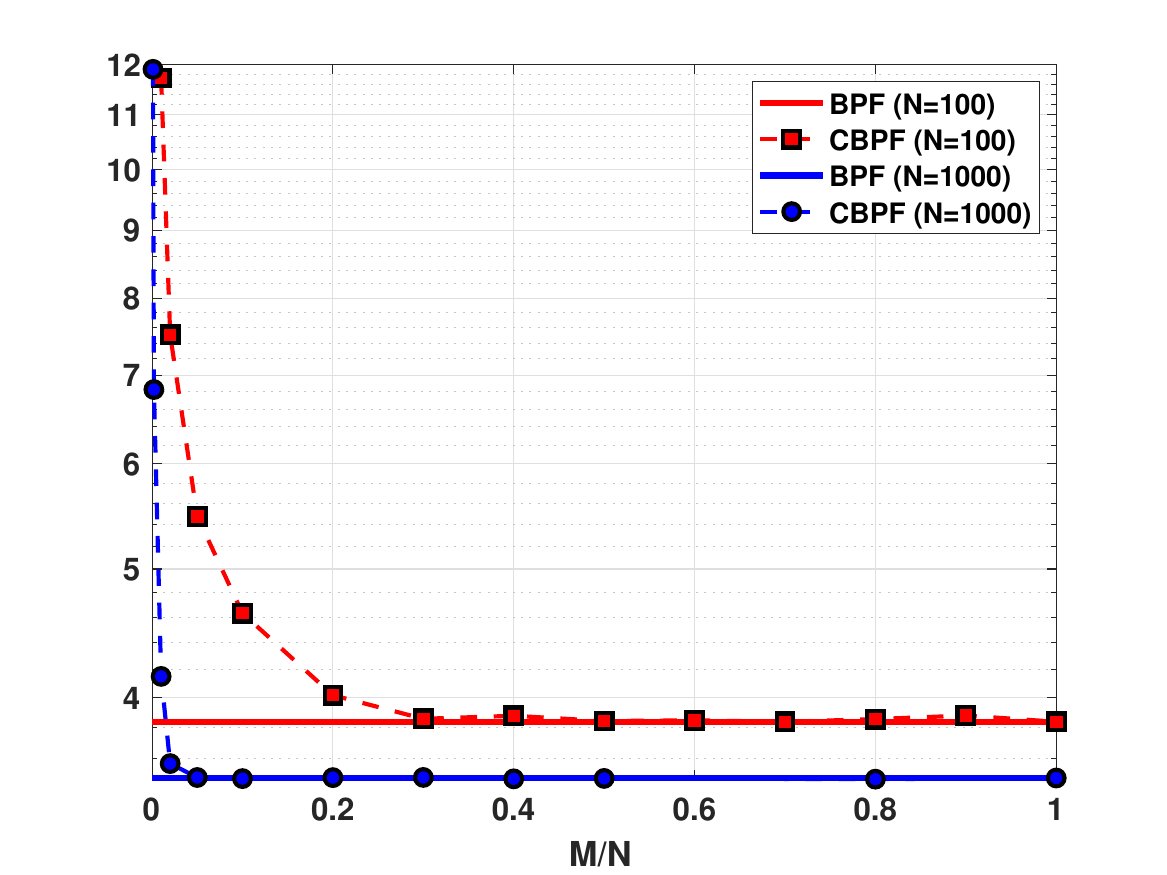}
\caption{RMSE (log-domain in the $y$-axis) as function of the ratio $\frac{M}{N}$. The dashed line with (squares and circles) corresponds to the CBPF with $N\in \{100,1000\}$. The solid lines correspond to the BPF with $N\in \{100,1000\}$. With $N=1000$,  CBPF  obtains the same performance of the BPF, with $98\%$ less evaluations of the likelihood function (i.e., only $M=20$ evalutions).}
\label{FigEX_CPF_2}
\end{figure}

Figure \ref{FigEX_CPF_2} shows the RMSE (averaged over $10^3$ independent runs) as function of the ratio $\frac{M}{N}$. The solid lines provide the RMSE obtained by the standard BPF. The dashed line with squares corresponds to the CBPF with $N=100$, whereas the dashed line with circles corresponds to the CBPF with $N=1000$.  Note that CBPF, with $N=100$,  obtains virtually the same performance of the BPF with $70\%$ less likelihood evaluations. With $N=1000$,   CBPF obtains the same performance of the BPF with $98\%$ less likelihood evaluations.
As in Section \ref{CPFsimu}, in this toy example, the likelihood evaluation is not expensive, and the differences in computational time are negligible.
 Fixing the compression rate $\frac{M}{N}$, we can also observe that the performance of CBPF improves when $N$ grows.

\begin{figure}[htbp]
\centering
\includegraphics[width=8cm]{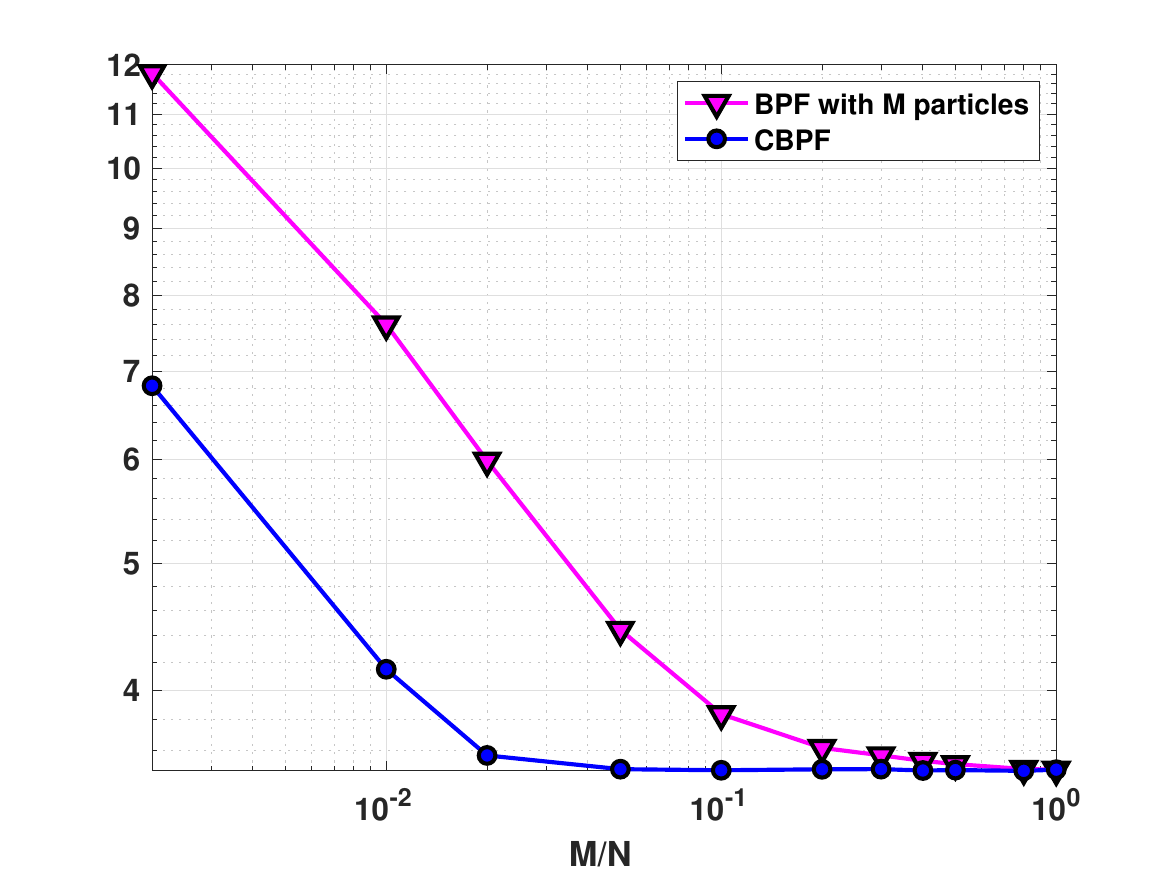}
\caption{RMSE (log-log-domain) as function of the ratio $\frac{M}{N}$. The circles correspond to the CBPF with different values of $M$ from $2$ to $1000$, and  $N=1000$. The triangles correspond to the standard BPF with $M$ particles. Both filters, CBPF and BPF, have the same number of likelihood evaluations.  CBPF always provides the smallest RMSE and the difference in RMSE increases when the compression is larger (left side of the figure). }
\label{FigEX_CPF_3}
\end{figure}

In Figure \ref{FigEX_CPF_3},  we compare the standard BPF and CBPF but, in this case, considering the same number of likelihood evaluations. Therefore, both the standard BPF and CBPF waste $M$ evaluations of the likelihood function per iteration.  We can observe that CBPF provides always the smallest RMSE and the difference in RMSE increases when the compression is bigger, i.e., in left side of the figure. Thus, fixing the likelihood evaluation budget, the proposed compression scheme  is an efficient procedure for managing this budget. The results in both Figures \ref{FigEX_CPF_2}-\ref{FigEX_CPF_3} show the benefits of the proposed approach.    


\subsection{Inference in Kepler's models}\label{KeplerSect}

In recent years, the problem of revealing objects orbiting other stars has acquired large attention. Different techniques have been proposed to discover exo-objects but, nowadays, the radial velocity technique is still the most used \cite{Gregory2011,Barros2016,Affer2019,Trifonov2019}. The problem consists in fitting a model (the so-called radial velocity curve) to data acquired at different moments spanning during long time periods (up to years). The model is highly non-linear and it is costly in terms of computation time (specially, for certain sets of parameters). Obtaining a value to compare to a single observation involves numerically integrating a differential equation in time or an iterative procedure for solving to a non-linear equation. Typically, the iteration is performed until a threshold is reached or $10^6$ iterations are performed. The problem of radial velocity curve fitting is applied in several related applications. It is similar to the problem of determining the orbits of spectroscopic binary stars \cite{Strassmeier1993,Galvez2006} or the stars surrounding the galactic center \cite{Gillessen2017}. In the following, we describe an orbital model, which is equivalent for any N-body system observed from Earth, i.e. exoplanetary systems, binary stellar system, double pulsars, etc. 

\begin{table}[t]
  \centering
  \caption{Description of parameters in Eq.~\eqref{eq:rv}.}
  \small
  \begin{tabular}{lll} 
   \hline
   Parameter & Description & Units \\
   \hline
   \multicolumn{3}{l}{For each planet}\\
   \hline
   $K_i$        & amplitude of the curve & m\,s$^{-1}$ \\
   ${u}_{i,t}$      & true anomaly     & rad \\
   $\omega_{i,t}$      & longitude of periastron & rad \\ 
   $e_i$        & orbit's eccentricity    & \ldots \\
   $P_i$        & orbital period        & s \\
   $\tau_i$       & time of periastron passage & s \\
   \hline
   \multicolumn{3}{l}{\footnotesize Below: not depending on the number of objects/satellite }\\
   \hline
   $V_0$      & mean radial velocity   & m\,s$^{-1}$ \\
   \hline
  \end{tabular}
  \label{tab:rvpar}
\end{table}

\subsubsection{Likelihood and transition functions}
When analysing radial velocity data of an exoplanetary system, it is commonly accepted that the \emph{wobbling} of the star around the centre of mass is caused by the sum of the gravitational force of each planet independently and that they do not interact with each other. Each planet follows a Keplerian orbit and the radial velocity of the host star is given by
\begin{equation}
{y}_{r,t} = V_0 + \sum\limits_{i = 1}^{S} K_i \left[ \cos \left( {u}_{i,t} + \omega_{i,t} \right) + e_i \cos \left( \omega_{i,t} \right) \right] +\xi_t,
\label{eq:rv}
\end{equation}
with $t=1,\ldots,T$ and $r=1,\ldots,R$. The number of objects in the system is $S$, that is consider known in this experiment (for the sake of simplicity). Both ${y}_{r,t}$, ${u}_{i,t}$ depend on time $t$, and then $\xi_t$ is a Gaussian noise perturbation with variance $\sigma_e^2$. For the sake of simplicity, we consider this value known, $\sigma_e^2=1$. 
The meaning of each parameter in Eq.~\eqref{eq:rv} is given in Table~\ref{tab:rvpar}. The likelihood
 function is defined by \eqref{eq:rv} and some indicator variables described below. 
 The angle ${u}_{i,t}$ is 
the true anomaly of the planet $i$ and it can be determined from
\begin{equation}
\frac{d{u}_{i,t}}{dt} = \frac{2\pi}{P_i} \frac{\left( 1 + e_i \cos {u_{i,t}} \right)^2}{\left( 1 - e_i \right)^\frac{3}{2}}
\label{eq:trueanomaly}
\end{equation}
As mentioned above, this equation has analytical solution. As a result, the true anomaly $u_t$ can be determined from the mean anomaly $M$. However, the analytical solution contains a non linear term that needs to be determined by iterating. First, we define the mean anomaly $M_{i,t}$ as
\begin{equation}
M_{i,t} = \frac{2\pi}{P_i} \left( t - \tau_i \right),
\label{eq:meananomaly}
\end{equation}
where $\tau_i$ is the time of periastron passage of the planet $i$ and $P_i$ is the period
of its orbit (see Table~\ref{tab:rvpar}). Then, through the Kepler's equation, 
\begin{equation}
M_{i,t} = E_{i,t} - e_i \sin E_{i,t},
\label{eq:kepler}
\end{equation}
where $E_{i,t}$ is the eccentric anomaly. Equation~\eqref{eq:kepler} has no analytic solution and it must be solved by an iterative procedure. A Newton-Raphson method is typically used to find the roots of this equation \textbf{\cite{Press2002}.} For certain sets of parameters this iterative procedure, can be particularly slow. We also have
\begin{equation}
\tan \frac{u_{i,t}}{2} = \sqrt{ \frac{1 + e_i}{1 - e_i}} \, \tan \frac{E_{i,t}}{2}, 
\label{eq:eccentricanomaly}
\end{equation}
%
The variables $\omega_{i,t}$'s, for $i=1,\ldots,N$, can vary with time. In particular, if the central body is much heavier than the other objects orbiting it or if the objects are very close, the so-called orbital precession is observed (e.g. \cite{Lin-Jia2013}). The state variable $\mathbf{x}$ is the vector
\begin{equation}
\mathbf{x}_t = [V_0, K_1, \omega_{1,t}, e_1, P_1, \tau_1, \ldots, K_S, \omega_{S,t}, e_S, P_S, \tau_S],
\end{equation}
For a single object (e.g., a planet or a natural satellite), the dimension of $\mathbf{x}_t$ is $d_X = 5+1=6$, with two objects the dimension of $\mathbf{x}_t$ is is $d_X = 11$ etc. Generally, we have $d_X=1+5S$. We also include in the likelihood function the  $V_0\in [-20,20]$, $K_{i} \in [0,50]$, $e_{i} \in [0,1]$, $P_i \in [0,365]$, $\omega_{i,t} \in [0,2\pi]$, $\tau_{i} \in [0,P_i]$ by means of indicator variables (i.e., the likelihood is zero outside these intervals), for all $i=1,\ldots,S$. This means that the likelihood function is zero when the particles fall out of these intervals. Note that the interval of $\tau_{i}$ is conditioned to the value $P_i$. This parameter is the time of periastron passage, i.e. the time passed since the object passed the closest point in its orbit. It has the same units of $P_i$ and can take values from 0 to $P_i$.
All the Eqs. \eqref{eq:rv}--\eqref{eq:eccentricanomaly}, jointly with the previous parameter constrains, induce a likelihood function
\begin{align*}
p({\bf y}_{1:T}|\mathbf{x}_{1:T})&= \prod_{t=1}^T p({\bf y}_t|\mathbf{x}_{t}),\\
&=\prod_{t=1}^T\prod_{r=1}^R p(y_{r,t}|\mathbf{x}_{t}).
\end{align*}
where ${\bf y}_t=[y_{1,t},\ldots,y_{R,t}]^{\top}$. Note that all the variables $\omega_{i,t}$ vary with the time, whereas the remaining components of ${\bf x}_t$ are static parameters. Sophisticated particle approaches could be used, for instance, combining MCMC and particle filtering schemes for addressing the inference of both dynamic and static parameters \cite{PMCMC,MARTINO2018134,GIS18}. The compressed particle idea can easily adapted to this scenario, or within more complicated particle algorithms. For the sake of simplicity, we leave it for future works.
Here, we use the simpler approach    
where we consider an artificial time-evolution of the parameters \cite{Liu01WESTb}, i.e., ${\bf x}_t = [V_{0,t}, K_{1,t}, \omega_{1,t}, e_{1,t}, P_{1,t}, \tau_{1,t}, \ldots, K_{S,t}, \omega_{S,t}, e_{S,t}, P_{S,t}, \tau_{S,t}]$, in order to have a prior transition equation for the entire state ${\bf x}_t$. We consider
\begin{equation}
{\omega}_{i,t}={\omega}_{i,t-1}+ v_t, \qquad t=1,\ldots,T,
\end{equation}
where $ v_t$ is a Gaussian noise perturbation with zero mean and variance $\sigma_V^2=0.5$. For the rest of parameters, we consider 
\begin{gather}
\left\{
\begin{split}
V_{0,t}&=V_{0,t-1}+ \xi_{0,t}, \\
K_{i,t}&=K_{i,t-1}+ \xi_{i,t}^{(1)}, \\
e_{i,t}&=e_{i,t-1}+ \xi_{i,t}^{(2)}, \\
P_{i,t}&=P_{i,t-1}+ \xi_{i,t}^{(3)}, \\
\tau_{i,t}&=\tau_{i,t-1}+ \xi_{i,t}^{(4)},
\end{split}
\right.
\end{gather} 
for $t=1,\ldots,T$ and $\xi_{0,t}$, $\xi_{i,t}^{(j)}$, $i=1,\ldots,S$, are Gaussian noises with zero mean and variance $\sigma_\xi^2=0.1$.
Thus, we have a Gaussian transition probability $p(\mathbf{x}_t|\mathbf{x}_{t-1})$. The initial probability $p({\bf x}_0)=\prod_{i=1}^{d_X} p(x_{i,0})$ is a product of marginal prior pdf, where $p(x_{i,0})$ is a uniform pdf for the parameter with constrains (see above) and, for the rest of parameters, $p(x_{i,0})$ is a Gaussian with zero mean and variance equal to $10$. The complete posterior is 
\begin{align*}
p(\mathbf{x}_{0:T}|{\bf y}_{1:T})=\frac{1}{p({\bf y}_{1:T})} p({\bf y}_{1:T}|\mathbf{x}_{1:T}) \left[\prod_{t=1}^T p(\mathbf{x}_t|\mathbf{x}_{t-1}) \right] p({\bf x}_0).
\end{align*}
where $Z=p({\bf y}_{1:T})$ is the complete marginal likelihood, that is also unknown. Note that to compute $Z=p({\bf y}_{1:T})$ we have to integrate out all the sequence of parameters $\mathbf{x}_{0:T}$ (trajectory), .i.e.,
\begin{align*}
p({\bf y}_{1:T})=\int_{\mathcal{X}^{T+1}} p({\bf y}_{1:T}|\mathbf{x}_{1:T}) \left[\prod_{t=1}^T p(\mathbf{x}_t|\mathbf{x}_{t-1}) \right] p({\bf x}_0) d\mathbf{x}_{0:T}. 
\end{align*}
We will approximate this integral via particle filtering (i.e., sequential importance sampling with resampling steps). See \cite{GIS18,GISssp16} for further details regarding the sequential estimation of $Z$.
\subsubsection{Inferring the number of objects orbiting the central mass}

Given a set of data $\{y_{r,t}\}_{r=1}^T$ for all time \textbf{instants} $t=1,\ldots,T$ generated according to the model (see the initial parameter values below), our goal is to infer the number of objects. For this purpose, we have to approximate the model evidence $Z=p({\bf y}_{1:T})$ via standard PF and the generic CPF in Table \ref{GCPFtable}. For a fair comparison, we consider the same ESS approximation, $\widehat{ESS}=\frac{1}{\sum_{n=1}^N (\bar{w}_t^{n})^2}$ and $\eta=0.5$.
 In all experiments, we set $R=5$ and $T=50$ and average the results over $500$ independent runs. 
 We consider three different experiments: {\bf (E1)} $S=0$, i.e., no object, {\bf (E2)} $S=1$ (one object) and {\bf (E2)} the case of two objects $S=2$. We set $V_0=2$, in all cases. For the first object in {\bf E1} and {\bf E2}, we set $K_{1,0}=25$, $\omega_{1,0}=0.61$, $e_{1,0}=0.1$, $P_{1,0}=15$, $\tau_{1,0}=3$. For {\bf E2}, we also consider a second object with $K_{2,0}=5$, $\omega_{2,0}=0.17$, $e_{2,0}=0.3$, $P_{2,0}=115$, $\tau_{2,0}=25$ (in that case $S=2$). Note that the SNR associate to the second object is low (so that the detection of this planet is not straightforward). The rest of trajectories are generated according to the transition model (and the corresponding measurements $y_{r,t}$ according to the observation model). 
  We consider $N=10^5$ total number of particles and just $M=100$ summary particles for CPF in Table \ref{GCPFtable} ($\frac{M}{N}=10^{-3}$). 

\subsubsection{Results}
At each run and for each experiment {\bf E1}-{\bf E2}-{\bf E3}, we run the particle filters considering different state dimensions and likelihood functions (according to Eq. \eqref{eq:rv}) computing $\widehat{Z}^{(1)}$ (corresponding to ``no planet''), $\widehat{Z}^{(2)}$ (corresponding to ``one planet'') and $\widehat{Z}^{(3)}$ (corresponding to ``two objects''). Then we obtain 
$$
j^*=\arg\max_{j^*\in\{1,2,3\}} \widehat{Z}^{(j)}.
$$
If $j^*=0$, we decide that there is no planet/satellite. If $j^*=1$, we decide that there is one object and if $j^*=2$ we assume that there are $2$ objects. The results are given in Tables \ref{E1}--\ref{E3}.
We compute the rate of each decision over the $500$ independent runs and for each scenario {\bf E1}-{\bf E2}-{\bf E3}.  Moreover, we provide the ``ranking'' of each decision, namely, how many times the specific decision has been the first choice, the second choice or the third choice. For instance, let us consider the decision ``zero object'' in Table \ref{E1}: the ranking in this case is $100-0-0$ which means that $100\%$ of cases the choice ``zero object'' have been the first one (i.e., with greater Bayesian evidence). In the same table, the ranking of the decision ``one object'' is $0-100-0$, i.e., this choice has been always the second possibility (with the second greater Bayesian evidence). 
 We can observe that, in {\bf E1}-{\bf E2}, we have no loss with CPF in term of detection, since we obtain the same results of the standard PF. However, CPF requires less computational time, saving almost the $70\%$ of the required time with the standard PF. In {\bf E3}, CPF decides more times ($67\%$) that there is only one object, which is an error since we have two objects in this scenario. However, also the standard PF decides $63\%$ of times ``one object''. In both cases, we always decide that there is at least one object (the choice ``zero object'' has been never selected).
Therefore, CPF provides very similar performance than a standard PF with much less computational cost.

 \begin{table}[h]
  \centering
  \caption{\small Experiment 1 {\bf E1} (no planet): percentage of the decisions (over $500$ runs) and the normalized computational time spent by each method.}
  \small
  \begin{tabular}{|c|c||c|c|c||c|} 
   \hline
  \multicolumn{2}{|c||}{Method}  & Zero & One & Two & Time \\
   \hline
    \hline
  \multirow{2}{*}{PF} & decision & 100$\%$ &0$\%$ & 0$\%$ &  \multirow{2}{*}{1}\\
    & ranking & {\scriptsize 100-0-0} &{\scriptsize 0-100-0} & {\scriptsize 0-0-100}& \\
    \hline
    \hline
   \multirow{2}{*}{CPF}& decision & 100$\%$ &0$\%$ & 0$\%$ & \multirow{2}{*}{0.32} \\
  & ranking & {\scriptsize 100-0-0} &{\scriptsize 0-100-0} & {\scriptsize 0-0-100}& \\
   \hline
  \end{tabular}
  \label{E1}
\end{table}

\begin{table}[h]
  \centering
  \caption{\small Experiment 2 {\bf E2} (one planet): percentage of the decisions (over $500$ runs) and the normalized computational time spent by each method.}
  \small
 \begin{tabular}{|c|c||c|c|c||c|} 
   \hline
  \multicolumn{2}{|c||}{Method}  & Zero & One & Two & Time \\
   \hline
    \hline
  \multirow{2}{*}{PF} & decision & 0$\%$ &88$\%$ & 12$\%$ &  \multirow{2}{*}{1}\\
    & ranking & {\scriptsize 0-0-100} &{\scriptsize 88-12-0} & {\scriptsize 12-88-0}& \\
    \hline
    \hline
   \multirow{2}{*}{CPF}& decision & 0$\%$ &88$\%$ & 12$\%$ & \multirow{2}{*}{0.32} \\
  & ranking & {\scriptsize 0-0-100} &{\scriptsize 88-12-0}& {\scriptsize 12-88-0}& \\
   \hline
  \end{tabular}
  \label{E2}
\end{table}

\begin{table}[h]
  \centering
  \caption{\small Experiment 3 {\bf E3} (two objects): percentage of the decisions (over $500$ runs) and the normalized computational time spent by each method.}
  \small
 \begin{tabular}{|c|c||c|c|c||c|} 
   \hline
  \multicolumn{2}{|c||}{Method}  & Zero & One & Two & Time \\
   \hline
    \hline
  \multirow{2}{*}{PF} & decision & 0$\%$ &64$\%$ & 36$\%$ &  \multirow{2}{*}{1}\\
    & ranking & {\scriptsize 0-0-100} &{\scriptsize 64-36-0} & {\scriptsize 36-64-0}& \\
    \hline
    \hline
   \multirow{2}{*}{CPF}& decision & 0$\%$ &67$\%$ & 33$\%$ & \multirow{2}{*}{0.32} \\
  & ranking & {\scriptsize 0-0-100} &{\scriptsize 67-33-0}& {\scriptsize 33-67-0}& \\
   \hline
  \end{tabular}
  \label{E3}
\end{table}



\subsection{PROSAIL inversion with time-varying physical parameters}\label{RemoteSect}
Earth observation from satellite sensors offers the possibility to monitor our planet with unprecedented accuracy. Radiative transfer models (RTMs) are forward models that encode the energy transfer through the atmosphere, and are used to model and understand the Earth system. These models also allow us to estimate the parameters that describe the status of the Earth from satellite observations by inverse modeling. However, performing inference over such simulators is generally an ill-posed problem because of the difficulty to invert the system and to compute the marginal likelihood.  Generally, RTMs are non-differentiable and computationally very costly models, which adds on a high level of difficulty in inference.

\begin{figure}[htbp]
\centering
\includegraphics[width=8cm]{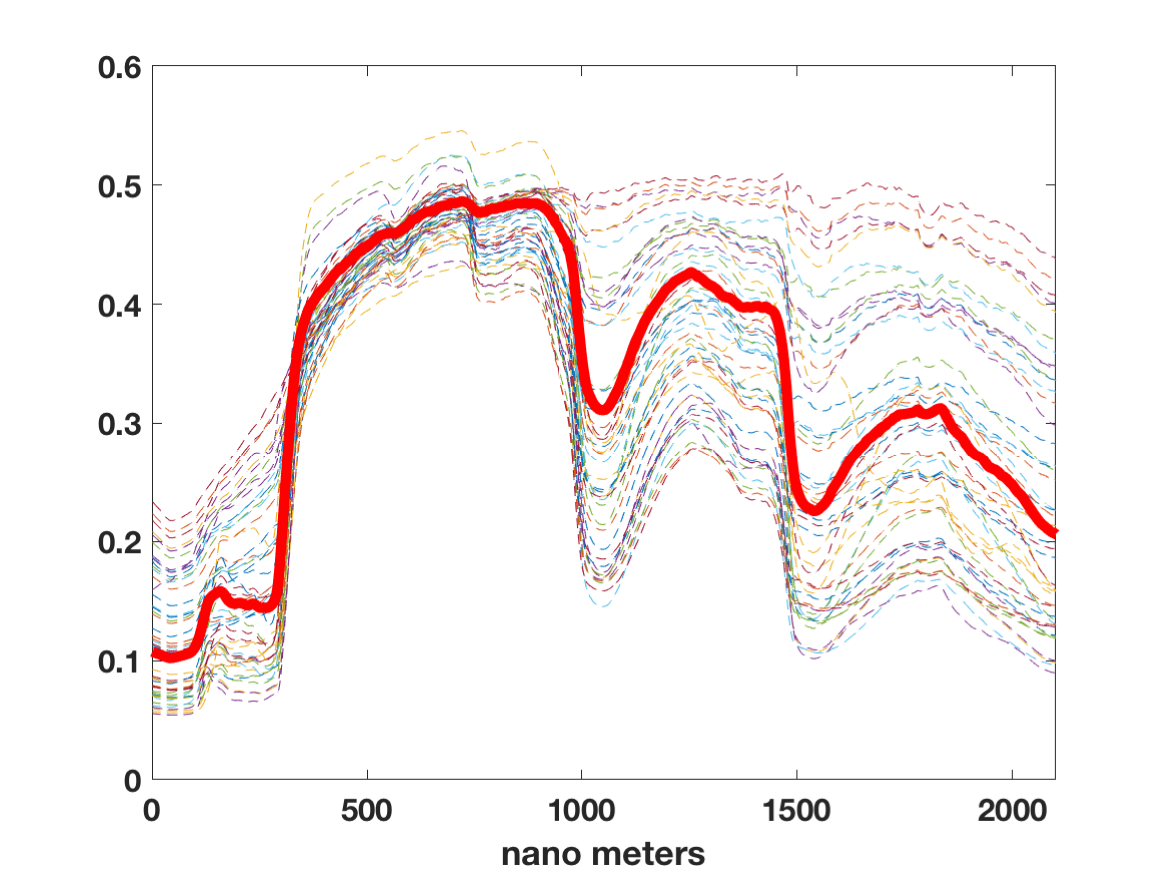}
\caption{The data ${\bf y}_t$ (solid line) and the model values corresponding to 50 particles, ${\bf f}^{(i)}={\bf f}({\bf x}_t^{(i)})$ (dashed lines), at $t=15$.}
\label{FigEX_prosail}
\end{figure}

Here we will test our method for inverting a commonly used radiative transfer model for vegetation monitoring. The so-called PROSAIL RTM is the most widely used model over the last two decades in remote sensing studies~\cite{jacquemoud2009prospect+}. It simulates reflectance as a function of:
\begin{itemize}
  \item[1)] A set of leaf optical properties, given by the mesophyll structural parameter (MSP), leaf chlorophyll (Chl), dry matter also referred as ``leaf mass per unit area'' (Cm), water (Cw), carotenoid (Car) and brown pigment (Cbr) contents.
  \item[2)] A set of canopy level characteristics, determined by leaf area index (LAI), the average leaf angle inclination (ALA) and the hot-spot parameter (Hotspot). System geometry is described by the solar zenith angle ($\theta_s$), view zenith angle $\theta_v$), and the relative azimuth angle between both angles ($\Delta\Theta$). 
  \end{itemize} 
In our experiments, we consider the inference of $7$ of these variables, that we also assuming varying in time. The rest of parameters are keep fixed to the default values in the PROSAIL code (\url{http://teledetection.ipgp.jussieu.fr/prosail/}), so that for simplicity they are assumed known. 
At time instant $t$, our state is 
$$
{\bf x}_t=[x_{1,t},x_{2,t},x_{3,t},x_{4,t},x_{5,t},x_{6,t},x_{7,t}]^{\top}
$$
where $x_{1,t}=\mbox{Chl}(t)$, $x_{2,t}=\mbox{Car}(t)$, $x_{3,t}=\mbox{Cbr}(t)$, $x_{4,t}=\mbox{Cw}(t)$, $x_{5,t}=\mbox{Cm}(t)$, $x_{6,t}=\mbox{MSP}(t)$, $x_{7,t}=\mbox{LAI}(t)$. The likelihood function at time $t$ is 
\begin{eqnarray}
 p({\bf y}_t|{\bf x}_t)=\exp\left(-\frac{1}{2\sigma^2}||{\bf y}_t-{\bf f}({\bf x}_t)||^2\right) \mathbb{I}_\mathcal{R}({\bf x}_t),
\end{eqnarray}
where ${\bf f}({\bf x}_t):\mathbb{R}^7\rightarrow \mathbb{R}^{2100}$ represents the PROSAIL model, and $\mathbb{I}_\mathcal{R}({\bf x}_t)$ is an indication function which is $1$ if ${\bf x}_t\in \mathcal{R}$ otherwise is $0$, if ${\bf x}_t\notin \mathcal{R}$. The region $\mathcal{R}$ is defined as $\mathcal{R}=\prod_{i=1}^7 \mathcal{I}_i$ with $\mathcal{I}_1=[0,100]$ ($\mu$g/cm$^2$), $\mathcal{I}_2=[0,25]$ ($\mu$g/cm$^2$),
 $ \mathcal{I}_3=[0,1]$, $ \mathcal{I}_4=[0,0.05]$ (cm),  $ \mathcal{I}_5=[0,0.02]$ (g/cm$^2$),  $ \mathcal{I}_6=[1,3]$, $\mathcal{I}_7=[0,1]$. The function ${\bf f}({\bf x}_t)=\mbox{PROSAIL}({\bf x}_t)$ is the high-nonlinear model represented by the code given at \url{http://teledetection.ipgp.jussieu.fr/prosail/}. The vector ${\bf y}_t \in \mathbb{R}^{2100}$ contains the measurements obtained by the satellite. The transition model is 
\begin{equation}
p({\bf x}_t|{\bf x}_{t-1})=\mathcal{N}({\bf x}_t|{\bf x}_{t-1},{\bm \Lambda}),
\end{equation}
where ${\bm \Lambda}$ is a diagonal $7 \times 7$ matrix with $\mbox{diag}[{\bm \Lambda}]=[1,0.4,10^{-2},10^{-3},10^{-3},0.4,0.4]^\top$. We recall that $t=1,\ldots,T$. We generate synthetic data $\{{\bf x}_{1:T},{\bf y}_{1:T}\}$ (setting $T=20$ and $\sigma^2=1$) according to the model starting with ${\bf x}_0=[40, 8, 0.2, 0.01,0.009,2.5,0.5]^{\top}$. We compare a standard PF with $N=10^4$ particles with a CPF with $M\in\{1000, 2000, 5000\}$ (and $N=10^4$). We also consider a standard PF with $N=M$ in order to show the benefits of the compression in CPF. 
 For all the filters we employ $\widehat{ESS}=\frac{1}{\sum_{n=1}^N (\bar{w}_t^{n})^2}$ with $\eta=0.5$.
 Figure \ref{FigEX_prosail} depicts the data ${\bf y}_t$ at $t=15$ (solid line) and the model values corresponding to 50 particles ${\bf f}^{(i)}={\bf f}({\bf x}_t^{(i)})$ (as an example).
 We compute the Root Mean Square Error (RMSE) in the estimation of the trajectory of parameters ${\bf x}_{1:T}$. The RMSE is obtained by averaging the square errors over all the component of the state, and over each time. We have averaged the results over $10^3$ independent runs. The results are given in Table \ref{PROSAIL}. We can observe that CPF provides very similar results than the standard PF with $N=10^4$ with much less computational cost. For instance, CPF with $M=10^3$ saves more than $95\%$ of the computational time with an increase of the MSE of only $5\%$. The comparison between CPF and the standard PFs with $N=M\in\{1000, 2000, 5000\}$ shows the benefit of the compression in CPF. 
 
\begin{table}[h]
  \centering
  \caption{\small Results of the PROSAIL inversion.}
  \small
 \begin{tabular}{|l||c|c|} 
   \hline
   Method & RMSE & Norm. TIME  \\
    \hline
     \hline
     Standard PF - $N=10^4$ &  3.99 & 1 \\
     \hline
     CPF - $M=10^3$ & 4.08 &  0.0112 \\
     CPF - $M=2\cdot 10^3$  & 4.04 & 0.1938\\
     CPF - $M=5\cdot 10^3$  & 3.99 & 0.4215 \\
     \hline
     Standard PF - $N=10^3$ &  4.30 & 0.0984  \\
     Standard PF - $N=2\cdot 10^3$ & 4.19& 0.1856  \\
     Standard PF - $N=5\cdot 10^3$ & 4.10 &0.4181 \\
     \hline

   \hline
  \end{tabular}
  \label{PROSAIL}
\end{table}
 


\section{Conclusions}
\label{ConSect}
We have introduced a novel approach for performing sequential Bayesian inference in the context of complex and costly models. In the proposed scheme, the expensive model is evaluated only in some well-chosen samples. The selection of these nodes is based on the so-called compressed Monte Carlo (CMC) scheme. The application of CMC within particle filtering schemes and the corresponding benefits are described and discussed. 
The provided theoretical and numerical results, which include applications in astronomy and remote sensing, showed the advantages of the proposed method.

{ \footnotesize
\section*{ \footnotesize Acknowledgements}
{LM acknowledges  support by  the Agencia Estatal de Investigaci{\'o}n AEI (project SPGRAPH, ref. num. PID2019-105032GB-I00) and the Found action by the Community of Madrid in the framework of the Multiannual Agreement with the Rey Juan Carlos University in line of action 1, ÒEncouragement of Young Phd students investigation" Project Ref. F661 Acronym Mapping-UCI.
JL-S acknowledges support by the Office of Naval Research (N00014-19-1-2226), Spanish Ministry of Science, Innovation and Universities (RTI2018-099655-B-I00) and Regional Ministry of Education and Research for the Community of Madrid (Y2018/TCS-4705). GCV was supported from the European Research Council (ERC) under the ERC Consolidator Grant 2014 project SEDAL (647423).
}


\bibliographystyle{IEEEtran}
\bibliography{bibliografia}

\appendix

\section{Proof of Theorem \ref{Teo2}}\label{appProof1}


First of all, we need to obtain some additional relationships. Let us define the partial estimators as 
\begin{equation}
{\widehat I}_m(h)=\sum_{i\in\mathcal{J}_m} {\bar w}_{m,i} h({\bf x}_i),
\end{equation}
an estimator of the integral 
\begin{equation}
\int_{\mathcal{X}_m} h({\bf x}) \bar{\pi}({\bf x}) d{\bf x}=\int_{\mathcal{X}} h({\bf x}) \bar{\pi}_m({\bf x}) d{\bf x},
\end{equation}
where we have set $\bar{\pi}_m({\bf x})= \bar{\pi}({\bf x}) \mathbb{I}(\mathcal{X}_m)$. with these definitions, note a that 
\begin{eqnarray}
{\widehat I}^{(N)}(h)=\sum_{i=1}^N {\bar w}_i h({\bf x}_i)&=&\sum_{m=1}^M \sum_{i\in\mathcal{J}_m} {\bar w}_i h({\bf x}_i), \nonumber \\
&=&\sum_{m=1}^M \widehat{a}_m \sum_{i\in\mathcal{J}_m} {\bar w}_{m,i} h({\bf x}_i) \nonumber\\
&=&\sum_{m=1}^M \widehat{a}_m {\widehat I}_m(h), \label{aquiComb_I}
\end{eqnarray}
where we have used ${\bar w}_{m,i}= \frac{{\bar w}_i}{\widehat{a}_m}$ as shown in Eq. \eqref{EqREVmagic}. Namely, the estimator ${\widehat I}^{(N)}(h)$ of $I(h)$ can be expressed as a convex combination of the $M$ partial estimators. A similar expression is valid for the particle approximations, i.e.,
\begin{eqnarray}
&&{\widehat \pi}^{(N)}({\bf x})=\sum_{m=1}^M \widehat{a}_m {\widehat \pi}_m({\bf x}), \quad \mbox{where} \\
&&{\widehat \pi}_m({\bf x})=\sum_{i\in\mathcal{J}_m} {\bar w}_{m,i} \delta ({\bf x}-{\bf x}_i).
\end{eqnarray}

{\it Proof.} Assume that $\mathcal{S}=\{{\bf x}_n,\bar{w}_n\}_{n=1}^N$ (hence also $N$) and the partition $\mathcal{P}$ are given and fixed (hence $M$ as well). Then, the summary weights $\widehat{a}_m$ are also fixed. The unique stochastic part in $ \widetilde{I}^{(M)}{(h)}$ is the selection of ${\bf s}_m$'s.
Let us consider the case when ${\bf s}_m$ is resampled randomly in each partition, according to the weights $ {\bar w}_{m,j}$ in Eq. \eqref{EqREVmagic}, i.e.,
$$
{\bf s}_m \sim {\widehat \pi}_m({\bf x}).
$$
 Given the set of weighted samples $\mathcal{S}=\{{\bf x}_n,{\bar w}_n\}_{n=1}^N$, note that 
\begin{eqnarray}
\mbox{E}[h({\bf s}_m)|\mathcal{S}]=\sum_{j\in \mathcal{J}_m} {\bar w}_{m,j} h({\bf x}_{j})=\widehat{I}_m(h).
\end{eqnarray}
Given Eq. \eqref{aquiComb_I}, we can also write 
\begin{eqnarray}\label{estaIMP}
\widehat{I}^{(N)}(h)=\sum_{m=1}^M \widehat{a}_m \widehat{I}_m(h)=\sum_{m=1}^M \widehat{a}_m \mbox{E}[h({\bf s}_m)|\mathcal{S}].
\end{eqnarray}
Note also that
\begin{eqnarray*}
\widetilde{I}^{(M)}(h)=\sum_{m=1}^M \widehat{a}_m h({\bf s}_m),
\end{eqnarray*}
Taking the expectation of both sides
\begin{eqnarray*}
\mbox{E}[\widetilde{I}^{(M)}(h)|\mathcal{S}]&=&\mbox{E}\left[\sum_{m=1}^M \widehat{a}_m h({\bf s}_m)\right] \\
&=&\sum_{m=1}^M \widehat{a}_m \mbox{E}[h({\bf s}_m)|\mathcal{S}],Ê\\
&=&\widehat{I}^{(N)}(h),
\end{eqnarray*}
where we have used Eq. \eqref{estaIMP}.
\section{Proof of Theorem \ref{Teo1}}
\label{App1}
Theorem \ref{Teo1} states that, with the choice $s_m=\sum_{j\in \mathcal{J}_m} {\bar w}_{m,j} h({\bf x}_j)$ in \eqref{EqSpecH}, we have ${\widetilde I}^{(M)}(f)\equiv {\widehat I}^{(N)}(h)$, for a specific function $h({\bf x})$ and $f({\bf x})={\bf x}$. Indeed, we have 
\begin{eqnarray}
{\widetilde I}^{(M)}(f)&=& \sum_{m=1}^M \widehat{a}_m s_m \nonumber \\
 &=& \sum_{m=1}^M \widehat{a}_m  \left[\sum_{j\in \mathcal{J}_m} {\bar w}_{m,j} h({\bf x}_j)\right], \nonumber 
 \end{eqnarray}
 where we have used $ s_m =\sum_{j\in \mathcal{J}_m} {\bar w}_{m,j} h({\bf x}_j)$. Replacing ${\bar w}_{m,j}$ with the expression ${\bar w}_{m,j}={\bar w}_j/\widehat{a}_m$ in Eq. \eqref{EqREVmagic}, we obtain a further simplification, 
 \begin{eqnarray}
{\widetilde I}^{(M)}(f)&=& \sum_{m=1}^M \widehat{a}_m  \left[\sum_{j\in \mathcal{J}_m} \frac{{\bar w}_j}{\widehat{a}_m} h({\bf x}_j)\right], \nonumber \\
&=& \sum_{m=1}^M \sum_{j\in \mathcal{J}_m} {\bar w}_j h({\bf x}_j)  \nonumber \\
&=& \sum_{j=1}^N {\bar w}_j h({\bf x}_{j})={\widehat I}^{(N)}(h), 
\end{eqnarray}
where we have also used the fact that, if we consider all the $M$ possible sums within $\mathcal{J}_m$, then we are considering all the possible $N$ samples and weights, i.e., $\sum_{m=1}^M \sum_{j\in \mathcal{J}_m} {\bar w}_j h({\bf x}_j)= \sum_{j=1}^N {\bar w}_j h({\bf x}_j)$.

\section*{Biographies}
\includegraphics[width=1in,height=1.25in,clip,keepaspectratio]{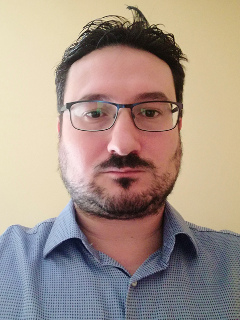}

{\bf Luca Martino} received his MSc in electronic engineering in 2006 at Politecnico di Milano. He obtained his PhD in Statistical Signal Processing from Universidad Carlos III de Madrid, Spain, in 2011. He was an Assistant Professor in the Department of Signal Theory and Communications at Universidad Carlos III de Madrid since then. In August 2013, he joined the Department of Mathematics and Statistics at the University of Helsinki. He worked as postdoctoral researcher also at the Universidade de S{\~a}o Paulo (USP) and at the Universitat de Valncia, Valncia, Spain. He has also been a visiting researcher at Universidade Federal do Rio de Janeiro (UFRJ). He is currently Professor with the Universidad Rey Juan Carlos de Madrid. His research interests include Bayesian inference, Monte Carlo methods and stochastic processes.
He holds a Hirsch's index h=28 (Google Scholar).

\vspace{2cm}
\includegraphics[width=1in,height=1.25in,clip,keepaspectratio]{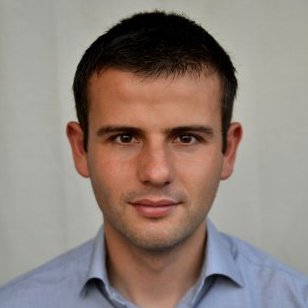}
 
 {\bf V\'ictor Elvira} is a Reader (Associate Professor) in Statistics and Data Science at the School of Mathematics at the University of Edinburgh (UK). He received his BSc and MSc in electrical engineering, and the PhD degree in statistical signal processing in 2011 from the University of Cantabria (Spain). From 2016 to 2019, he was an Associate Professor at the engineering school IMT Lille Douai (France). From 2013 to 2016, he was an Assistant Professor at University Carlos III of Madrid. He has also been a visiting researcher at several institutions such as Stony Brook University of New York (USA) and Paris-Dauphine University (France). Dr. Elvira's research interests are mostly in the fields of statistical signal processing, computational statistics, and machine learning, in particular in Monte Carlo methods for Bayesian inference with different applications including sensor networks, wireless communications, target tracking, ecology, and biomedicine. He is a Turing Fellow, Fulbright Fellow, and a Marie Curie Fellow. He is an Associate Editor of the IEEE Transactions on Signal Processing.  

\vspace{0.5cm}

\includegraphics[width=1in,height=1.25in,clip,keepaspectratio]{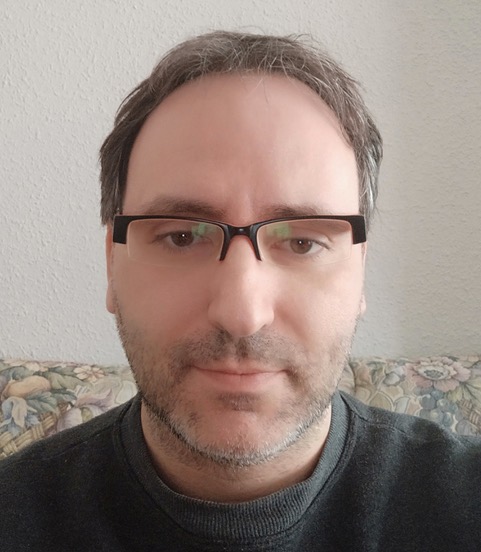}

{\bf Javier Lopez-Santiago} has a tenure track position in Signal Theory and Communications at the Polytechnic School at the University Carlos III of Madrid (Spain). He received his PhD degree in Physics from the Universidad Complutense de Madrid in 2005 and obtained the Extraordinary Price to the best PhD thesis in Physics at the UCM that year. From 2005 to 2015, he devoted himself exclusively to research in Astrophysics. In 2005, he became a Marie Curie Fellow at the Istituto Nazionale di Astrofisica in Italy. Between 2007 and 2016, Dr. Lopez-Santiago worked at several research institutes such as the Astronomical Observatory of Palermo in Italy, the Physics Faculty of the Universidad Complutense de Madrid and the World Space Observatory Ground Segment in Spain. In 2019, he received the UC3M social Council Excellence Award to young researchers. Dr. Lopez-Santiago's research interests are Bayesian Inference Methods, Machine Learning techniques applied to physical and biological processes, Wavelet Analysis applied to oscillatory phenomena in the Sun, Space Debris collision risk and re-entry probabilities, among others. He is member of the Royal Spanish Physics Society.

\vspace{1cm}
\includegraphics[width=1in,height=1.25in,clip,keepaspectratio]{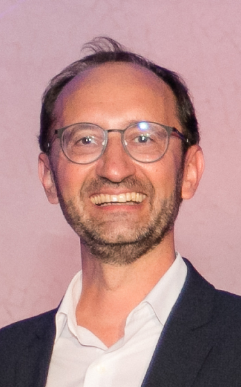}

{\bf Gustau Camps-Valls} (IEEE Fellow'18, IEEE Distinguished lecturer, PhD in Physics) is currently a Full professor in Electrical Engineering and head of the Image and Signal Processing (ISP) group, http://isp.uv.es. He is interested in the development of machine learning algorithms for geosciences and remote sensing data analysis. He is an author of around 250 journal papers, more than 300 conference papers, 20 international book chapters, and editor of 6 books on kernel methods and deep learning. He holds a Hirsch's index h=72 (Google Scholar), entered the ISI list of Highly Cited Researchers in 2011, and Thomson Reuters ScienceWatch identified one of his papers on kernel-based analysis of hyperspectral images as a Fast Moving Front research. He received two European Research Council (ERC) grants: an ERC Consolidator grant on "Statistical learning for Earth observation data analysis" (2015) and an ERC Synergy grant on "Understanding and Modelling the Earth system with machine learning" (2019). In 2016 he was included in the prestigious IEEE Distinguished Lecturer program of the GRSS.

\end{document}